\documentclass[11pt]{article}

\usepackage{amsmath, amsfonts, amssymb,latexsym,wasysym,graphicx}
\input epsf.sty

\newtheorem{Lemme}{Lemme}[section]
\newtheorem{Proposition}[Lemme]{Proposition}

\newtheorem{Definition}[Lemme]{D\'efinition}

\newcommand{\BEQ}{\begin{equation}}     % Gleichungen Anfang ..
\newcommand{\BEA}{\begin{eqnarray}}
\newcommand{\BD}{\begin{displaymath}}
\newcommand{\EEQ}{\end{equation}}       % .. und Ende
\newcommand{\EEA}{\end{eqnarray}}
\newcommand{\ED}{\end{displaymath}}
\newcommand{\del}{\delta}
\newcommand{\Del}{\Delta}
\newcommand{\eps}{\varepsilon}          % epsilon

\newcommand{\g}{{\mathfrak{g}}}

\renewcommand{\Vert}{{\mathrm{Vert}}}
\newcommand{\Hor}{{\mathrm{Hor}}}
\newcommand{\R}{\mathbb{R}}
\newcommand{\C}{\mathbb{C}}
\newcommand{\Z}{\mathbb{Z}}
\newcommand{\N}{\mathbb{N}}
\newcommand{\D}{\mathbb{D}}
\renewcommand{\P}{\mathbb{P}}

\newcommand{\Id}{{\mathrm{Id}}}

\def\proba{{\mathbb{P}}}
\def\esper{{\mathbb{E}}}
\def\T{{\mathbb{T}}}
\def\F{{\mathbb{F}}}
\def\Var{{\mathrm{Var}}}

\newcommand{\eop}{\hfill $\Box$}        % quod erat demonstrandum ...

\newcommand{\II}{{\rm i}}               % gerades i fuer komplexe Einheit
\renewcommand{\Re}{{\rm Re\ }}          % Realteil
\newcommand{\half}{{1\over 2}}          % 1/2 als Bruch
\renewcommand{\vec}[1]{\boldsymbol{#1}} % Vektoren fettgedruckt
\catcode`\@=11
\def\numberbysection{\@addtoreset{equation}{section}
        \def\theequation{\thesection.\arabic{equation}}}
\numberbysection

\begin{document}

\vspace*{1.5cm}
\begin{center}
{\Large \bf Mode d'emploi de la th\'eorie constructive des champs bosoniques}

{\bf avec une application aux chemins rugueux}

\end{center}

\vspace{2mm}
\begin{center}
{\bf  J\'er\'emie Unterberger}
\end{center}

\vspace{2mm}
\begin{quote}

\renewcommand{\baselinestretch}{1.0}
\footnotesize
{Nous d\'eveloppons dans cet article les principaux arguments constructifs utilis\'es en th\'eorie quantique des champs, en nous cantonnant aux th\'eories bosoniques, pour lesquelles
il n'existe pas de pr\'esentation g\'en\'erale r\'ecente. L'article s'adresse d'abord et avant tout \`a des math\'ematiciens ou physiciens math\'ematiciens connaissant les arguments de base
de la th\'eorie perturbative des champs, et souhaitant conna\^\i tre un cadre g\'en\'eral dans lequel ils peuvent \^etre rendus rigoureux. Il fournit \'egalement un aper\c cu d'une s\'erie d'articles
r\'ecents \cite{MagUnt1,MagUnt2} visant \`a donner une d\'efinition constructive des chemins rugueux et du calcul stochastique fractionnaire.

\medskip
We develop in this article the principal constructive arguments used in quantum field theory, limiting us to bosonic theories, for which there does not exist any recent general
presentation. The article is primarily written for mathematicians or mathematical physicists knowing the basic arguments of quantum field theory, and desiring to discover a general
framework in which they can be made rigorous. It also provides a glimpse of a recent series of articles \cite{MagUnt1,MagUnt2} whose aim is to give a constructive definition of
rough paths and fractionary stochastic calculus.
}
\end{quote}

\vspace{4mm}
\noindent
{\bf Mots cl\'e:}
 th\'eorie constructive des champs, renormalisation, d\'eveloppements cluster, Brownien fractionnaire, calcul stochastique fractionnaire, chemins rugueux.
 


\tableofcontents

%\medskip

\section{Introduction}

Cet article est n\'e d'un expos\'e inachev\'e, fait au s\'eminaire de physique math\'ematique de l'Universit\'e de Lyon en f\'evrier 2011. L'auteur de ces lignes voulait raconter  bri\`evement ses travaux
r\'ecents sur les chemins rugueux et sur le calcul stochastique fractionnaire \cite{MagUnt1,MagUnt2}, avant d'en venir \`a l'objet essentiel de sa conf\'erence -- celui pour lequel il avait \'et\'e mandat\'e, pourrait-on
dire --, \`a savoir des explications d\'etaill\'ees concernant les m\'ethodes constructives en th\'eorie des champs, sur lesquelles reposent de mani\`ere essentielle ses r\'esultats - ainsi
que, plus g\'en\'eralement, la validit\'e des calculs perturbatifs tournant autour de la renormalisation. De fil en
aiguille, d'explication en explication, et faute de combattants (pourtant tr\`es patients!) l'expos\'e a d\^u s'interrompre au bout de presque trois heures.

Les {\em m\'ethodes constructives}  sont en g\'en\'eral m\'econnues du grand public (int\'eress\'e \`a la th\'eorie des champs sous ses divers aspects, s'entend), et r\'eput\'ees r\'eserv\'ees \`a une poign\'ee d'experts. Il y
a l\`a un paradoxe, compte tenu du succ\`es extraordinaire de la th\'eorie des champs et des th\'eories de renormalisation. Celles-ci sont devenues incontournables en  physique des particules de haute \'energie comme en physique
de la mati\`ere condens\'ee, depuis les travaux de K. Wilson dans les ann\'ees 70. Parall\`element, le champ de leurs applications math\'ematiques n'a cess\'e de s'\'etendre ces vingt derni\`eres ann\'ees; on pourrait citer, en suivant approximativement l'ordre
chronologique, les invariants de Donaldson pour les vari\'et\'es de dimension 4 et la th\'eorie de Seiberg-Witten \cite{Moo};  les int\'egrales matricielles et leurs
applications \`a la combinatoire \cite{LanZvo}; la th\'eorie des noeuds et la topologie en basse dimension \cite{Baez}; la quantification par d\'eformation
des vari\'et\'es de Poisson par M. Kontsevich \cite{Kon}; la d\'emonstration constructive du th\'eor\`eme de Kolmogorov-Arnold-Moser (KAM) \cite{GalKAM, BGK}; les travaux fondateurs d'A. Connes et D. Kreimer portant sur la reformulation alg\'ebrique de l'algorithme de renormalisation de
Bogolioubov, Parasiuk, Hepp et Zimmermann (BPHZ), \`a l'origine d'une multitude de d\'eveloppements plus formels, avec des applications notamment aux sch\'emas num\'eriques d'int\'egration
des \'equations diff\'erentielles \cite{CK,Bro}; et bien d'autres encore. 

Des raisons existent \`a cela, de bonnes et de mauvaises. Passons sur le fait que les applications d'essence combinatoire ou formelle -- en somme, la majorit\'e des applications aux
math\'ematiques, lorsqu'on laisse de c\^ot\'e l'analyse et les probabilit\'es -- ne s'int\'eressent par d\'efinition qu'aux d\'eveloppements asymptotiques, sans se soucier de leur
convergence. La principale et plus s\'erieuse est sans doute la limitation (qu'on peut esp\'erer provisoire!)  du champ d'application des
m\'ethodes constructives aux th\'eories qui sont des perturbations de th\'eories  gaussiennes. Les th\'eories physiques sous-jacentes s'\'ecrivent en termes d'un lagrangien d'interaction
${\cal L}_{int}$ coupl\'e \`a un param\`etre $\lambda>0$, appel\'e constante de couplage; la th\'eorie est gaussienne \`a la limite $\lambda=0$.  Comme nous l'expliquons plus loin (cf. sections 1 et 2),
la renormalisation (perturbative comme constructive) repose sur une int\'egration successive sur les degr\'es de libert\'e de la th\'eorie, en commen\c cant par les \'echelles d'\'energies (ou moments de Fourier) les plus hautes (ce que les physiciens appellent la zone {\em ultra-violette}) et en descendant jusqu'aux \'energies les plus basses (dans l'{\em infra-rouge}). Ce faisant, les param\`etres de la th\'eorie,
notamment $\lambda$, sont renormalis\'es, et deviennent d\'ependants de l'\'echelle. La th\'eorie effective \`a l'\'echelle $j$, donnant le comportement des fonctions \`a $n$ points $\langle
\psi(x_1)\ldots\psi(x_n)\rangle$, avec $\log |x_i-x_{i'}|\simeq -j$ \footnote{ou encore (notant $\psi^j$ les fluctuations gaussiennes du champ $\psi$ sur une distance typique de l'ordre
de $2^{-j}$), des fonctions $\langle\psi^j(x_1)\ldots\psi^j(x_n)\rangle$, sans restriction sur les $x_i$.}, se calcule alors comme une perturbation finie de la th\'eorie gaussienne
avec les param\`etres effectifs \`a l'\'echelle $j$. Tout le probl\`eme de la renormalisation provient de ce que le flot du groupe de renormalisation associ\'e 
(sauf dans les th\'eories dites {\em asymptotiquement libres} pour lesquelles la constante de couplage effective tend vers $0$ \footnote{La chromodynamique quantique est asymptotiquement libre
dans l'ultra-violet \cite{LeB}; au contraire, le mod\`ele $\phi^4$ non massif en dimension 4 \'etudi\'e ici est asymptotiquement libre dans l'infra-rouge.}, et dans certaines th\'eories avec des sym\'etries particuli\`eres pour lesquelles le flot est trivial \cite{RVTW06}) fait cro\^itre le param\`etre $\lambda$
au-del\`a de la zone o\`u les perturbations sont permises. Dans certains cas \footnote{On peut citer la th\'eorie $\phi^4$ non massive en dimension 2 et 3, associ\'e au mod\`ele d'Ising \`a la temp\'erature
critique.}, des d\'eveloppements perturbatifs en diagrammes de Feynman \`a 2 ou 3 boucles (au-del\`a l'utilisation de l'ordinateur devient indispensable) semblent
indiquer l'existence d'un {\em point fixe non trivial}, autrement dit d'une valeur limite {\em non nulle} du param\`etre $\lambda$ renormalis\'e aux \'energies les plus basses, signature
d'un comportement effectif non gaussien de la th\'eorie \`a l'\'echelle macroscopique. Dans d'autres, de mani\`ere encore plus critique -- comme l'\'electrodynamique quantique en dimension 4,
suppos\'ee d\'ecrire l'interaction des \'electrons avec le champ \'electromagn\'etique, autrement dit de la lumi\`ere et de la mati\`ere --, ces m\^emes d\'eveloppements perturbatifs en diagrammes
de Feynmann sugg\`erent que la constante de couplage effective cro\^it sans limite dans la limite ultra-violette, et donc que le mod\`ele est intrins\`equement mal d\'efini aux \'energies les plus hautes.  Malheureusement ces mod\`eles
sont parmi les plus fondamentaux en physique de la mati\`ere condens\'ee comme en physique des particules de haute \'energie. La majorit\'e des physiciens s'est donc tourn\'ee vers la
recherche de nouveaux mod\`eles (comme la grande unification ou la th\'eorie des cordes), ou des mod\`eles int\'egrables (comme c'est le cas du mod\`ele d'Ising et de nombre d'autres mod\`eles
de physique statistique en dimension 2), ou encore a abandonn\'e l'id\'ee de d\'eveloppements perturbatifs rigoureux, se contentant de d\'eveloppements perturbatifs en diagrammes de Feynman.

Un point de terminologie important s'impose ici avant de poursuivre: on appelle d'ordinaire {\em th\'eorie perturbative des champs} un d\'eveloppement en diagrammes de Feynman. Le d\'eveloppement
asymptotique g\'en\'eral de la fonction de partition ou des fonctions \`a $n$ points ou corr\'elations de la th\'eorie fait appara\^itre une s\'erie enti\`ere en $\lambda$ ou en $\hbar$, notoirement
divergente,  dans toutes les th\'eories connues; plus pr\'ecis\'ement, le coefficient de $\lambda^n$, ou somme des diagrammes de Feynman \`a $N$ boucles$^*$, diverge comme une {\em factorielle}
$N!$ \`a une certaine puissance. Pourtant, les mod\`eles explor\'es en {\em th\'eorie constructive des champs} sont \'egalement {\em perturbatifs} \`a leur mani\`ere, puisqu'ils ne sont bien
d\'efinis -- comme nous venons de l'expliquer -- que lorsque les param\`etres renormalis\'es restent {\em petits} \`a toutes les \'echelles. La cl\'e r\'eside dans un d\'eveloppement perturbatif contr\^ol\'e --
une sorte de d\'eveloppement de Taylor avec reste int\'egral, au lieu d'un d\'eveloppement en s\'erie -- permettant d'exprimer la fonction de partition, ou plut\^ot son logarithme, l'\'energie libre,
comme la limite d'une s\'erie de fonctions $f_j$ analytiques en $\lambda$,  $\sum_{j=-\infty}^{+\infty} f_j(\lambda)$ portant sur les \'echelles d'\'energies, de l'ultra-violet
$(j=+\infty)$ jusqu'\`a l'infra-rouge $(j=-\infty)$. Formellement, le d\'eveloppement en s\'erie (divergent  \`a nouveau) des $f_j(\lambda)$ permet de retrouver la s\'erie des diagrammes de Feynman;
les d\'eveloppements perturbatifs de la th\'eorie constructive peuvent donc se voir comme une resommation partielle astucieuse de cette m\^eme s\'erie. 
La correspondance entre la s\'erie divergente des diagrammes de Feyman et la s\'erie convergente de la th\'eorie constructive est en fait plus pr\'ecise que cela. L'\'energie libre est typiquement analytique \`a l'int\'erieur d'un petit disque $\{\lambda\in \C; |\lambda-\lambda_0|<\lambda_0\}$ $(\lambda_0>0)$ inclu dans le demi-plan $\Re\lambda>0$, et des estim\'ees permettent
de montrer qu'elle est \'egale \`a la somme de Borel de son d\'eveloppement en s\'erie en $0$, donn\'e par les diagrammes de Feynman.

Apr\`es ces pr\'ecautions d'usage, nous pouvons maintenant tenter de r\'esumer les succ\`es \`a porter au cr\'edit de la th\'eorie constructive; on pourra se reporter \`a la monographie
de V. Rivasseau \cite{Riv}, \`a la th\`ese de A. Abdesselam \cite{Abd}, ou aux livres de V. Mastropietro et de M. Salmhofer \cite{Mas,Sal} (qui s'int\'eressent plus sp\'ecifiquement aux th\'eories fermioniques).
La th\'eorie
constructive des champs est un programme lanc\'e \`a l'origine dans les ann\'ees 60 par 
A. S. Wightman   \cite{Wigh}, dont le but \'etait de donner des exemples explicites de th\'eories des champs avec une interaction non triviale; cf. \cite{GJb} 
et les  r\'ef\'erences donn\'ees dans l'article pour une  bibliographie plus \'etendue. E. Nelson  donna la premi\`ere contribution au programme  en 1965 en introduisant une
analyse multi-\'echelles \cite{Nel} afin de contr\^oler la divergence du mod\`ele $\phi^4$ en deux dimensions, dont la seule divergence provient de l'ordre de Wick. 
J. Glimm et A. Jaffe    introduisirent l'analyse g\'en\'erale sur l'espace des phases \cite{GJa} pour des mod\`eles avec un nombre fini de graphes  divergents. Le d\'eveloppement
en clusters fut invent\'e par   J. Glimm, A. Jaffe et T. Spencer \cite{GJS} pour contr\^oler la limite en volume infini.

L'\'ecole romaine \cite{Benfa} se rendit alors compte que cette analyse de l'espace des phases \'etait en un certain sens une version spatiale continue du {\em d\'eveloppement
en blocs de spins} ou {\em block-spin expansion}, \'ecrite en premier par  Kadanoff pour le mod\`ele d'Ising, et devenue ult\'erieurement un outil majeur \`a la fois
en physique des particules de haute \'energie et en physique statistique  gr\^ace \`a l'introduction  par K. Wilson du concept de {\em groupe de renormalisation} \cite{Wilb,Wilc}.
L'outil du d\'eveloppement   multi-\'echelles  fut d\'evelopp\'e  dans les ann\'ees 80 afin de donner une version rigoureuse de groupe de renormalisation de Wilson, en introduisant notamment
le flot des param\`etres effectifs (ou renormalis\'es), cf. 
\cite{GKc} pour l'approche \`a la fa\c con blocs de spins, et  \cite{FMRS}  pour la version continue appel\'ee {\em d\'eveloppement en clusters multi-\'echelle} ou {\em multi-scale cluster expansion}.

Cette derni\`ere approche a \'et\'e poursuivie durant les trente derni\`eres ann\'ees, conduisant \`a la fois \`a des avanc\'ees conceptuelles
 \cite{Abd,AbdRiv1,AbdRiv2,MagRiv}, et \`a des 
applications \`a des mod\`eles de la th\'eorie quantique des champs avec un comportement asymptotique non trivial soit dans la limite {\em ultra-violette} (ou des hautes \'energies, ou encore
\`a petite distance)  ou {\em infra-rouge} (ou des basses \'energies, ou \`a grande distance); on peut citer p\^ele m\^ele  le mod\`ele $\phi^4$-model, l' \'electrodynamique
ou la   chromodynamique quantique, les mod\`eles de physique statistique sur r\'esau, les marches al\'eatoires auto-\'evitantes
\cite{MI}...  D'autres  r\'egimes avec une  singularit\'e non plus ponctuelle (dans la limite $|\xi|\to\infty$ en moments de Fourier), mais au voisinage d'une {\em  surface} interpr\'et\'ee
comme  surface de  Fermi, ont \'egalement \'et\'e \'etudi\'es, avec des  applications \`a la localisation d' Anderson et \`a des mod\`eles de  fermions non-relativistes en    interaction, tels que le mod\`ele de  Luttinger  \cite{Benfb} en dimension un ou le mod\`ele du jellium en dimension deux  \cite{FMRT}, en relation avec la th\'eorie de  Bardeen-Cooper-Schrieffer (BCS) sur la
   supraconductivit\'e. En g\'en\'eral, les th\'eories  fermioniques peuvent \^etre trait\'ees sans introduire tout le lourd appareillage des d\'eveloppements en clusters multi-\'echelles  \cite{AbdRiv3}. 
Malgr\'e les avanc\'ees conceptuelles de la derni\`ere d\'ecennie, l'avis (personnel) de l'auteur est que le cadre g\'en\'eral pour ces d\'eveloppements en clusters, tel qu'on le trouve dans l'article
 \cite{FMRS} \'ecrit il y a plus de vingt ans, \'etait  le plus appropri\'e, \`a la fois par sa g\'en\'eralit\'e, et son usage parcimonieux d'identit\'es combinatoires et d'alg\`ebre, au profit de
 d\'eveloppements en arbres somme toute intuitifs et faciles \`a visualiser, dans lesquels apparaissent clairement les id\'ees essentielles. Le d\'efaut majeur (mais partag\'e dans une
 large mesure par les autres approches) est la technicit\'e assez redoutable
 des bornes finales; celle-ci n'est pas apparente dans l'article -- bien qu'il contienne tous les arguments majeurs --, mais r\'eelle. Nous esp\'erons que cet article peut servir
 de compagnon \`a l'article \cite{MagUnt2}, dans lequel le lecteur pourra trouver tous les d\'etails dans une pr\'esentation nouvelle.

\bigskip

Pr\'esentons maintenant bri\`evement le long chemin qui, partant de probl\`emes fondamentaux concernant la d\'efinition m\^eme du calcul stochastique fractionnaire, nous a conduits \`a chercher des r\'eponses en th\'eorie des champs -- et plus pr\'ecis\'ement
en th\'eorie {\em constructive} des champs, puisqu'il s'agit ici d'analyse et de probabilit\'es.

L'\'etude des \'equations diff\'erentielles stochastiques dirig\'ees par le brownien (ou, via la formule d'It\^o ou celle de Feyman-Kac, des \'equations de diffusion) 
est un des th\`emes essentiels de la th\'eorie des  probabilit\'es, depuis les premiers travaux d'Einstein et Smoluchowski. L'int\'er\^et port\'e \`a cette th\'eorie tient \`a ce qu'elle
se confond dans une large mesure avec celle des processus de Markov continus. L'outil technique essentiel pour le calcul stochastique est la th\'eorie des (semi-)martingales,
qui repose elle-m\^eme sur le caract\`ere Markovien du processus, ainsi que sur la notion de {\em variation quadratique} des trajectoires, finie lorsque celles-ci sont de  r\'egularit\'e
H\"older d'indice $\ge 1/2$. Il est bien connu que deux th\'eories d'int\'egration (celle d'It\^o et celle de Stratonovich) sont en concurrence. Celle de Stratonovich est sans doute celle qui
correspond le plus \`a l'intuition, puisqu'elle s'obtient comme limite de l'int\'egrale de Riemann usuelle contre de bonnes approximations $C^1$ par morceaux des trajectoires
browniennes (notamment les classiques interpolations lin\'eaires par morceaux), et qu'elle v\'erifie la formule fondamentale du calcul infinit\'esimal,  \`a savoir
  $F(B(t))=F(B(s))+\int_s^t F'(B(u)) d^{Strato}B(u)$ pour toute fonction r\'eguli\`ere $F$ \'evalu\'ee le long d'une trajectoire brownienne $B(t)$ (cf. \cite{WZ}, cit\'e dans \cite{KS}, \S 5.2 D).

Cette approche \'echoue lorsqu'on consid\`ere des processus stochastiques d'indice de r\'egularit\'e H\"older $\alpha<1/2$ \footnote{Rappelons qu'un chemin continu
  $X:[0,T]\to\R$ est $\alpha$-H\"older, $\alpha\in(0,1)$, si $\sup_{s,t\in[0,T]} \frac{|X_t-X_s|}{|t-s|^{\alpha}}<\infty$. Les trajectoires browniennes sont $\alpha$-H\"older pour
  tout $\alpha<1/2$; elles ont une variation quadratique finie presque s\^urement en raison de compensations al\'eatoires.}.
Le champ d'applications est immense et mal d\'efini, allant des diffusions sur des fractals \cite{HL} aux sous-diffusions en milieu poreux \cite{Goldenfeld,Lesne}, des processus multi-fractionnaires
(et leurs proches parents, les marches al\'eatoires multi-fractales  ou martingales multiplicatives \cite{BM} avec leurs applications en gravit\'e quantique de Liouville \cite{DS} ou pour mod\'eliser
la turbulence \cite{Frisch,FalGawVer,KupMur}) aux {\em bruits color\'es}, utilis\'es de mani\`ere ph\'enomonologique dans nombre d'applications, en particulier en synth\`ese d'images.  Ce sont ces derniers qui nous int\'eresseront ici, en raison de leur
caract\`ere gaussien qui les apparente aux mod\`eles usuels de la th\'eorie quantique des champs. L'exemple le plus \'el\'ementaire de ces processus gaussiens est appel\'e 
habituellement {\em brownien fractionnaire}$^*$ \footnote{Ils s'obtiennent
en effet comme d\'eriv\'ee ou int\'egrale fractionnaire -- suivant la valeur de $\alpha$ -- du brownien.} par les probabilistes; il s'agit en fait d'une famille de processus index\'ee
par un param\`etre $\alpha$ correspondant \`a l'indice de r\'egularit\'e H\"older des trajectoires \cite{PLV,Nua}.  Or les travaux de L. Coutin et Z. Qian \cite{CQ02}
ont montr\'e que l'int\'egrale stochastique non triviale la plus \'el\'ementaire construite \`a partir du brownien fractionnaire bidimensionnel $\phi=(\phi_1(t),\phi_2(t))$ --
avec ses deux composantes ind\'ependantes et de m\^eme loi --, \`a savoir

\BEQ {\cal A}(s,t):=\int_s^t d\phi_1(t_1)\int_s^{t_1} d\phi_2(t_2)=\int_s^t (\phi_2(u)-\phi_2(s))d\phi_1(u) , \EEQ
une int\'egrale it\'er\'ee d'ordre $2$, d\'efinie comme limite des int\'egrales it\'er\'ees des interpolations lin\'eaires par morceaux des trajectoires,  {\em diverge} quand
$\alpha\le 1/4$. Des travaux ult\'erieurs reposant sur des m\'ethodes diff\'erentes \cite{CQ02,Nua,Unt08,Unt08b} ont confirm\'e l'existence de cette barri\`ere apparemment
infranchissable en $\alpha=1/4$.

Et pourtant, la th\'eorie des {\em chemins rugueux$^*$} (ou {\em rough paths}), une th\'eorie d'int\'egration adapt\'ee aux chemins irr\'eguliers, introduite par T. Lyons \`a la fin des ann\'ees 90 \cite{Lyo98,LyoQia02} et
devenue un outil essentiel en calcul stochastique  \cite{Lyo98,LyoQia02,Gub,Lej,Lej-bis,FV}, pr\'edit -- malheureusement par des arguments g\'eom\'etriques non constructifs -- l'existence
d'approximations $C^1$ par morceaux, autres que l'interpolation lin\'eaire par morceaux, dont les int\'egrales it\'er\'ees de tous ordres convergent vers des quantit\'es finies s'interpr\'etant comme
{\em substituts d'int\'egrales it\'er\'ees du brownien fractionnaire} -- en termes plus g\'eom\'etriques, comme {\em chemin rugueux} au-dessus du brownien --. Nous insistons sur l'id\'ee qu'il
s'agit de substituts: A. Lejay \cite{Lej,Lej-bis} a bien fait voir comment on peut modifier \`a loisir les int\'egrales it\'er\'ees d'un chemin en ins\'erant tout le long des "bulles" microscopiques
invisibles \`a l'oeil nu. Les travaux de l'auteur \cite{Unt-ren,Unt-fBm,Unt-Holder,Unt-resume,MagUnt1,MagUnt2} ont montr\'e en fait que lesdites int\'egrales it\'er\'ees s'expriment en termes de champs singuliers (dits {\em ordonn\'es
en Fourier}) qui peuvent \^etre r\'egularis\'es par l'ajout d'un terme d'interaction dans le lagrangien, sans modifier les trajectoires du champ r\'egulier $\phi$ sous-jacent. Ce miracle
s'explique par le flot quasi-trivial du groupe de renormalisation, {\em une} it\'eration suffisant pour \'ecranter totalement l'interaction. On obtient ainsi toute une classe
de processus gaussiens "g\'en\'eralis\'es"   qu'on pourrait appeler
 {\em champs quasi-gaussiens}. La th\'eorie g\'en\'erale \cite{FoiUnt} donne une multitude d'int\'egrales it\'er\'ees possibles correspondant \`a des choix essentiellement arbitraires de champs
 singuliers; la construction physique, quant \`a elle, donne une famille \`a un param\`etre d'int\'egrales it\'er\'ees construites par un proc\'ed\'e naturel en th\'eorie des champs, pouvant de plus  s'interpr\'eter comme ajout
 d'une d\'erive (drift) singuli\`ere dans une \'equation diff\'erentielle stochastique  \cite{MagUnt3}, et permettant de d\'efinir une int\'egration stochastique avec de bonnes propri\'et\'es, adapt\'ee
 par exemple \`a la r\'esolution d'\'equations diff\'erentielles stochastiques.  

\bigskip

Les termes fran\c cais suivis d'une ast\'erisque $( \ ^*\ )$ sont retraduits en anglais dans le lexique final, afin de permettre au lecteur anglophone de comprendre ais\'ement, et au lecteur
francophone de se r\'ef\'erer aux articles originaux. La th\'eorie constructive n'ayant jamais \'et\'e \'ecrite en fran\c cais, leur traduction
fran\c caise a \'et\'e obtenue en suivant les traditions orales, pour lesquelles l'auteur remercie chaleureusement Jacques Magnen, sans qui -- plus g\'en\'eralement  -- tout ce travail de transcription
de la th\'eorie constructive n'aurait pas \'et\'e possible.

%%%%%%%%%%%%%%%%%%%%%

\section{Comptage de puissance pour les diagrammes de Feynman multi-\'echelles}

%%%%%%%%%%%%%%%%%%%%%%%%%%%%%%%

Curieusement, les trait\'es classiques sur la th\'eorie quantique des champs (qu'ils introduisent l'algorithme de Bogolioubov-Parasiuk-Hepp-Zimmermann ou qu'ils en restent \`a des
d\'eveloppement \`a 1 ou 2 boucles$^*$) n'utilisent pas la notion du  {\em comptage de puissance$^*$}, ni de {\em diagramme de Feynman multi-\'echelles$^*$}. C'est pourtant (et les sp\'ecialistes
le savent bien depuis longtemps \cite{FMRS1,FMRS2,GalNic}) de loin le moyen le plus simple de montrer qu'on peut renormaliser les diagrammes de Feynman de fa\c con \`a produire des
quantit\'es finies, et aussi de borner ces diagrammes. Nous renvoyons ici au livre de V. Rivasseau \cite{Riv}, ou \`a la th\`ese r\'ecente de F. Vignes-Tourneret \cite{FVT} o\`u ces estim\'ees
classiques sont red\'emontr\'ees en d\'etails de mani\`ere tr\`es p\'edagogique.

\medskip
Le point de d\'epart est une th\'eorie gaussienne. Soit donc $\psi:\R^D\to\R^d$ un champ gaussien stationnaire sur $\R^D$ \`a $d$ composantes, de noyau de covariance $C_{\psi}(x,y)=C_{\psi}(x-y)$.
 Dans toute la suite on choisit une d\'ecomposition en \'echelles $M$-adique du champ, o\`u $M$ est une constante $>1$ fix\'ee.
 
 \begin{Definition}
 \begin{enumerate}
\item Soit $\chi^0:\R^D\to\R$, resp. $\chi^1$ une fonction $\ge 0$ \`a support compact telle que $\chi^0\equiv 1$ dans un voisinage de $0$, resp. $\chi^1\equiv 0$ dans un voisinage de $0$ et
 $\chi^1\equiv 1$ dans un voisinage du bord de l'hypercube d\'efini par $\sup_{j=1,\ldots,D} |\xi_j|=1$. Ces deux fonctions peuvent \^etre choisies de sorte que $(\chi^0,(\chi^j)_{j\ge 1})$,
 avec $\chi^j:=\chi^1(M^{-j}\cdot)$, d\'efinissent une partition de l'unit\'e, i.e. $\chi^0+\sum_{j\ge 1}\chi^j\equiv 1$. Soit $\rho\ge 0$. Alors la {\em troncature$^*$ ultra-violette
 \`a l'\'echelle $\rho$} d'une fonction $f:\R^D\to\R^d$ est $f^{\to\rho}:={\cal F}^{-1}\left(\xi\mapsto \left[\sum_{j=0}^{\rho} \chi^j(\xi)\right] {\cal F}f(\xi)\right)$, o\`u $\cal F$
 d\'esigne la transformation de Fourier. En termes simples, la troncature ultra-violette "coupe" toutes les composantes Fourier de moment $\xi$ tel que $|\xi|\gg M^{\rho}$. 
 \item Posons $C_{\psi}^j:={\cal F}^{-1}\left(\xi\mapsto \chi^j(\xi){\cal F}C_{\phi}(\xi)\right)$. Alors $\psi$ a la m\^eme loi que la s\'erie de champs gaussiens ind\'ependants
 $\sum_{j\ge 0}\psi^j$, si $\psi^j$ a pour covariance $C_{\psi}^j$. La troncature$^*$ ultra-violette d'\'echelle $\rho$ du champ $\psi$ est $\psi^{\to\rho}:=\sum_{j=0}^{\rho} \psi^j$, de
 covariance $C_{\psi}^{\to\rho}:=\sum_{j=0}^{\rho} C_{\psi}^j$.
 \end{enumerate}
 \end{Definition}
 
 Le cas typique est celui d'un {\em champ gaussien multi-\'echelles de dimension $\beta$}, pour lequel $|{\cal F}C_{\psi}(\xi)|\approx \frac{1}{|\xi|^{D-2\beta}}$ et $|\langle \psi^j(x)\psi^j(y)\rangle|\le C_r
 \frac{|x-y|^{-2\beta}}{(1+M^j|x-y|)^r}$ pour tout $r\ge 1$ (les exposants $-(D-2\beta)$ et $-2\beta$ se correspondant par transformation de Fourier), cf. \cite{MagUnt1}.  La d\'efinition de la dimension
 $\beta$ suit la d\'efinition usuelle des  physiciens, en ce sens que $\psi(x)$ se comporte comme une \'energie (ou inverse de distance) \`a la puissance $\beta$ \footnote{La convention dans \cite{MagUnt1}
 est d'appeler dimension d'\'echelle$^*$ $-\beta$, \'egal \`a la r\'egularit\'e H\"older des trajectoires.}. En particulier, les champs de dimension d'\'echelle $\beta>0$ sont divergents dans
 l'ultra-violet.
 
 \begin{Definition}[champs en interaction]
 Soit $\vec{\lambda}:=(\lambda_1,\ldots,\lambda_q)\in\C^q$ un ensemble de param\`etres, et $P_1,\ldots,P_q\ (q\ge 1)$ des polyn\^omes homog\`enes sur $\R^d\to(\R^d)^D$. Alors la {\em th\'eorie
 en interaction} avec {\em lagrangien d'interaction} ${\cal L}_{int}(\psi)(x):=\sum_{p=1}^q \lambda_p P_p(\psi(x);\nabla\psi(x))$ est (si elle existe!) la limite faible $\proba_{\vec{\lambda}}(d\psi)$
 des mesures de Gibbs
 \BEQ \proba_{\vec{\lambda},V,\rho}(d\psi):=\frac{1}{Z_{\vec{\lambda},V,\rho}} e^{-\int_V {\cal L}_{int}(\psi^{\to\rho})(x)dx} d\mu^{\to\rho}(\psi\big|_V),\EEQ
 quand le volume $|V|$ et l'\'echelle de cut-off ultra-violet $\rho$ tendent vers l'infini, o\`u: $V\subset\R^D$ est compact; $d\mu^{\to\rho}(\psi\big|_V)$ est la mesure gaussienne
 correspondant au champ r\'egularis\'e $\psi^{\to\rho}$ restreint au volume fini $V$; $Z_{\vec{\lambda},V,\rho}$ est une constante de normalisation appel\'ee {\em fonction de partition}.
 \end{Definition}
 
 En d\'eveloppant l'exponentielle de l'interaction en s\'erie, et en utilisant la formule de Wick bien connue (rappel\'ee en \S 3.2), on peut exprimer les fonctions \`a $n$ points (ou corr\'elations) de la
 th\'eorie, $\langle \psi_{i_1}(x_1)\ldots\psi_{i_n}(x_n)\rangle_{\vec{\lambda}}$, comme une somme formelle de diagrammes de Feynman, $\frac{1}{Z_{\vec{\lambda}}} \sum_{\Gamma}
 A(\Gamma)$, o\`u $\Gamma$ parcourt l'ensemble des diagrammes avec $n$ lignes externes $\psi_{i_1}(x_1),\ldots,\psi_{i_n}(x_n)$, et $A(\Gamma)\in\C$ est l'\'evaluation du diagramme correspondant;
 les fonctions \`a $n$ points {\em connexes} s'obtiennent alors comme somme sur les diagrammes {\em connexes}, sans la constante de renormalisation $\frac{1}{Z_{\vec{\lambda}}}$. Si l'on
 choisit une \'echelle pour chaque ligne (int\'erieure ou ext\'erieure) du graphe, on obtient ce qu'on peut appeler un {\em diagramme de Feynman multi-\'echelles}.
 
 Dans toute la suite on supposera pour simplifier le comptage de puissances que la th\'eorie est {\em juste renormalisable}; autrement dit, les constantes de couplage $\lambda_i$ sont
 sans dimension, ou encore, chaque terme dans l'interaction $\int_{\R^D} \psi_{i_1}(x)\ldots\psi_{i_I}(x)dx$ est de dimension $\beta_{i_1}+\ldots+\beta_{i_I}-D=0$. On d\'emontre alors
 facilement qu'un diagramme mono-\'echelle $\Gamma$ dont toutes les lignes int\'erieures et ext\'erieures sont d'une m\^eme \'echelle fix\'ee $j$ est d'ordre $M^{j\omega(\Gamma)}$, o\`u
 $\omega(\Gamma):=D-\sum_{\ell\in L_{ext}(\Gamma)} \beta_{i_l}$ ($L_{ext}(\Gamma)$ d\'esignant l'ensemble des lignes externes) est le {\em degr\'e de divergence superficielle} du graphe.
 Pour la th\'eorie $\phi^4$ en 4 dimensions par exemple, le champ $\phi$ est de dimension $\beta=\frac{D}{2}-1=1$, et le degr\'e de divergence superficielle d'un graphe \`a $n$ lignes externes est
 $4-n$ (cf. \cite{LeB}, \S 6.1.2).   Un diagramme mono-\'echelle est bien entendu convergent. Les divergences en th\'eorie quantique des champs proviennent des {\em diagrammes quasi-locaux},
 c'est-\`a-dire des diagrammes multi-\'echelles $\Gamma$ dont les lignes internes sont toutes de plus haute \'energie que les lignes externes; on note $i_{\Gamma}$ l'\'echelle {\em minimum}
 des lignes {\em internes}, et $e_{\Gamma}$ l'\'echelle {\em maximum} des lignes {\em externes}, de sorte que la {\em hauteur$^*$} $ht_{\Gamma}:=i_{\Gamma}-e_{\Gamma}$ d'un diagramme quasi-local $\Gamma$ est positive. En raison de la d\'ecroissance polynomiale ou exponentielle
 "scal\'ee" de la covariance $C^j(x-y)=\langle \psi^j(x)\psi^j(y)\rangle$, n\'egligeable sur des distances $|x-y|\gg M^{-j}$, l'ordre de grandeur de l'\'evaluation $A(\Gamma)$ d'un diagramme quasi-local s'obtient en int\'egrant sur des vertex int\'erieurs  \`a distance $\lesssim M^{-i_{\Gamma}}$ les uns des autres, inf\'erieure \`a la distance a priori comparable \`a $M^{-e_{\Gamma}}$ des
 vertex ext\'erieurs, d'o\`u le nom "quasi-local". Dans la terminologie de \cite{Riv,FVT}, les diagrammes quasi-locaux {\em divergents} sont dits {\em diagrammes dangereux}
 \footnote{La formule des arbres de Zimmermann (ou algorithme de renormalisation de BPHZ) se r\'e\'ecrit dans ce langage multi-\'echelles, faisant appara\^\i tre essentiellement des soustractions
 de contre-termes associ\'es \`a des for\^ets de diagrammes dangereux, et rendant quasi-imm\'ediate la preuve de la finitude des graphes renormalis\'es (cf. \cite{FVT},
 \S 1.3.3).}.  Pour \'evaluer un diagramme multi-\'echelles, on r\'ealise une sorte d'"\'ecorch\'e": pour chaque \'echelle $j$ des lignes du diagramme, on dessine l'ensemble des lignes
 d'\'echelle $\ge j$, not\'e $\Gamma^{j\to}$. Soit $j_{J}>j_{j-1}>\ldots$ la liste d\'ecroissante des \'echelles du diagramme. On construit un arbre couvrant de $\Gamma$ en extrayant du graphe $G^{j_J\to}$ un arbre couvrant,
 puis en le compl\'etant en un arbre couvrant de $G^{j_{J-1}\to}$, et ainsi de suite jusqu'\`a \'epuisement des \'echelles. On int\`egre successivement sur les sommets de l'arbre couvrant en partant
 des \'echelles les plus hautes. La d\'ecroissance "scal\'ee" des propagateurs montre que chaque vertex contribue un facteur de l'ordre de 1 (pour une th\'eorie juste renormalisable bien
 entendu). En int\'egrant sur tous les sommets de l'\'echelle $j_J$ sauf un, on trouve une contribution totale major\'ee par $M^{Dj_J}\cdot M^{-ht_{\Gamma} \sum_{\ell
  \in L_{ext}(\Gamma^{j_J\to})}  \beta_{j_{\ell}}}$. En d\'ecomposant $ht_{\Gamma}$ en $j_J-j_{J-1}$, on trouve finalement $M^{j_J}$ \`a un exposant $\omega(\Gamma^{j_J\to})=D-\sum_{\ell
   \in L_{ext}(\Gamma^{j_J\to})}  \beta_{j_{\ell}}$. Si tous les degr\'es de divergence superficiels $\omega(\Gamma^{j_i\to})$ de tous les sous-diagrammes quasi-locaux sont $<0$, on obtient
   ainsi des {\em facteurs de ressort$^*$} permettant de sommer sur les diff\'erences d'\'echelle $j_i-j_{i-1}$ et de d\'emontrer que le diagramme de Feynman habituel (obtenu en sommant
   sur toutes les attributions d'\'echelle possibles) est convergent. La {\em renormalisation} consiste \`a soustraire l'\'evaluation des sous-diagrammes quasi-locaux superficiellement divergents \`a moments
   externes nuls (ou mieux encore son d\'eveloppement de Taylor \`a l'ordre $\tau$ autour des moments externes nuls), ce qui est \'equivalent \`a d\'eplacer toutes les lignes externes au m\^eme point. Le graphe \'etant quasi-local, c'est-\`a-dire {\em quasi-ponctuel} du point de vue
   de l'\'echelle de ses lignes externes, on comprend que cette op\'eration donne la contribution principale du graphe. La soustraction est \'equivalente du point de vue du comptage de puissance
   \`a remplacer $\omega(\Gamma^{j_i\to})$ par $\omega^*(\Gamma^{j_i\to}):=\omega(\Gamma^{j_i\to})-\tau-1$. Pour $\tau$ suffisamment grand ($\tau=2$ pour la th\'eorie $\phi^4$ en dimension 4, $\tau=0$ pour le mod\`ele
   de chemins rugueux),   $\omega^*(\Gamma^{j_i\to})<0$, ce qui permet de sommer sur les \'echelles.

%%%%%%%%%%%%%%%%%%%%%%%%%%%%%%%%%%%%%%%%%%%%%%
%%%%%%%%%%%%%%%%%%%%%%%%%%%%%%%%%%%%%%
%%%%%%%%%%%%%%%%%%%%%%%%%%%%%%%%%%%%\`u

\section{D\'eveloppement en clusters}

Les d\'eveloppements en clusters$^*$  proviennent de l'\'etude de mod\`eles sur r\'eseau ou de gaz dilu\'es \`a haute temp\'erature \cite{ItzDro,Rue}. Dans tous les cas, l'id\'ee est d'\'evaluer
une fonction de partition portant sur un grand nombre de degr\'es de libert\'e coupl\'es avec une interaction \`a courte port\'ee en la r\'e\'ecrivant comme somme sur des amas finis  ("clusters") 
enti\`erement d\'ecoupl\'es. Les d\'eveloppements en clusters  de la th\'eorie des champs sont fond\'ees sur une d\'ecomposition en ondelettes simplifi\'ee $(\psi^j_{\Del^j})$
d'un champ gaussien $\psi$, o\`u $j$ est une \'echelle de Fourier {\em verticale} comme pr\'ec\'edemment,  et $\Del^j$ un cube {\em horizontal} (i.e. relatif \`a l'espace direct $\R^D$ et non \`a
l'espace de Fourier) de taille $M^{-j}$ autour du centre de la composante ondelette. Chaque $\psi^j_{\Del^j}$ peut \^etre vu comme un {\em degr\'e de libert\'e $^*$} de la th\'eorie, 
ces diff\'erents degr\'es de libert\'e \'etant relativement ind\'ependants les uns des autres, de sorte que l'interaction  int\'egr\'ee $\int {\cal L}_{int}(\psi)(x)dx$ peut se r\'e\'ecrire comme une double s\'erie horizontale et verticale,
divergente horizontalement (en raison de l'invariance par translation de la th\'eorie) et verticalement (sauf si la th\'eorie ne n\'ecessite pas de renormalisation). Les d\'eveloppements
en clusters horizontaux (H) et verticaux (V) permettent de r\'e\'ecrire la fonction de partition $Z_V^{\to\rho}$ sur un volume fini, avec \'echelle de troncature ultra-violette $\rho$, comme
une somme,
\BEQ Z_V^{\to\rho}=\sum_n \frac{1}{n!} \sum_{\P_1,\ldots,\P_n \ {\mathrm{non-overlapping}}} F_{HV}(\P_1)\ldots F_{HV}(\P_n), \label{eq:ZVo} \EEQ
o\`u:

-- $\P_1,\ldots,\P_n$ sont des {\em polym\`eres} disjoints, i.e. des ensembles de cubes $\Del$ reli\'es par des liens horizontaux et verticaux; pendant les d\'eveloppements en clusters,
la mesure gaussienne a \'et\'e modifi\'ee de sorte que les composantes des champs appartenant \`a des polym\`eres diff\'erents sont devenues ind\'ependantes;

-- $F_{HV}(\P)$, $\P=\P_1,\ldots,\P_n$ est l'\'evaluation $\langle f_{HV}(\P)\rangle_{\lambda}$ d'une fonction $f_{HV}$ d\'ependant uniquement des composantes situ\'ees dans le support de $\P$.

Les faits fondamentaux sont les suivants: (i) la {\em fonction d'\'evaluation du polym\`ere$^*$} $F_{HV}(\P)$ est d'autant plus petite que le polym\`ere est grand, en raison de la d\'ecroissance
polynomiale ou exponentielle \`a grandes distances (pour la direction horizontale), et par des arguments de comptage de puissance pour la direction verticale, ce qui conduit \`a l'image d'\^ilots horizontaux$^*$ maintenus
ensemble par des {\em ressorts$^*$} verticaux; (ii) les liens horizontaux et verticaux dans $\P$ (une fois qu'{\em un} cube appartenant \`a $\P$ a \'et\'e fix\'e) suppriment l'invariance
par translation, responsable de la divergence lorsque $|V|\to\infty$. Une astuce combinatoire classique, appel\'ee {\em d\'eveloppement de Mayer} (un d\'eveloppement en clusters d'un type
particulier en fait) permet de r\'e\'ecrire  l'\'eq. (\ref{eq:ZVo}) comme une somme similaire sur des {\em arbres de polym\`eres$^*$}, parfois appel\'es {\em Mayer-extended polymers}
(polym\`eres \'etendus \`a la Mayer ?) et not\'es $\P$ eux aussi par abus de notation, mais sans contrainte de non-overlap, $Z_V^{\to\rho}=\sum_n \frac{1}{n!}
\sum_{\P_1,\ldots,\P_n} F(\P_1)\ldots F(\P_n)$, o\`u $F=F_{HVM}$ est une nouvelle fonction d'\'evaluation de polym\`ere tenant compte du d\'eveloppement de Mayer, de sorte que
$\log Z_V^{\to\rho}=\sum_{\P} F(\P)$ appara\^it comme une quantit\'e extensive puisqu'invariante par translation. Les choses sont en fait plus compliqu\'ees que cela, les consid\'erations
pr\'ec\'edentes ne valant que pour une \'echelle donn\'ee. Au total, on trouve que, dans la limite $|V|,\rho\to\infty$, l'\'energie
libre $\log Z_V^{\to\rho}$ est une somme sur chaque \'echelle de quantit\'es extensives d\'ependant de l'\'echelle consid\'er\'ee, i.e. $\log Z_V^{\to\rho}=|V|\sum_{j=0}^{\to\rho} M^{Dj}
f_V^{j\to\rho}$, o\`u $f_V^{j\to\rho}$ (qu'on peut voir comme une \'energie libre par cube d'\'echelle $j$) converge quand $|V|\to\infty$ vers une quantit\'e finie de l'ordre de $O(\lambda)$. On retrouve l'id\'ee que chaque cube d'\'echelle $j$ contient un
degr\'e de libert\'e. Alternativement, on peut approcher cette d\'ecomposition par une somme $\sum_{j=0}^{\rho} \int_V f_{vol}^{j\to\rho}(x)dx$, o\`u $f_{vol}^{j\to\rho}$ est une
{\em densit\'e volumique} d'\'energie libre d'\'echelle $j$, de l'ordre de $M^{Dj}$. Finalement, les fonctions \`a $n$ points se calculent de la m\^eme mani\`ere en incorporant des champs externes d'une \'echelle donn\'ee et en sommant sur les \'echelles de ces champs.

%%%%%%%%%%%%%%%%%%%%%%%%

\subsection{Formule de Brydges-Kennedy-Abdesselam-Rivasseau}

Voyons maintenant comment concr\'etiser ces id\'ees. La tr\`es jolie formule ci-dessous, obtenue dans une premi\`ere version par Brydges et Kennedy \cite{BK} (cf. toutefois
\cite{Erice} pour des versions encore ant\'erieures), puis am\'elior\'ee et syst\'ematis\'ee
par A. Abdesselam et V. Rivasseau \cite{AbdRiv1,AbdRiv2}, fascine les math\'ematiciens \cite{Car}. Elle permet de traiter le {\em d\'eveloppement en clusters horizontal} ainsi que le {\em 
d\'eveloppement de Mayer};
dans le premier cas, les {\em objets} sont des {\em cubes} d'une \'echelle donn\'ee, dans le deuxi\`eme cas, des {\em polym\`eres multi-\'etages}.

Cette formule s'\'enonce de mani\`ere abstraite sur un ensemble fini d'objets $\cal O$ quelconque. Un lien $\ell$ de l'ensemble $\cal O$ est une paire ${o_{\ell},o'_{\ell}}$ d'objets
distincts. L'ensemble des liens de $\cal O$ est not\'e $L({\cal O})$; on peut le voir comme l'ensemble des ar\^etes du graphe total sur $\cal O$. Si $\F$ est une for\^et (autrement dit,
une r\'eunion disjointe d'arbres) reliant les objets de l'ensemble $\cal O$, on note  $L(\F)\subset  L({\cal O})$ l'ensemble de ses ar\^etes.

\begin{Definition}[formule de Brydges-Kennedy-Abdesselam-Rivasseau]

Soit $Z_{\cal O}:[0,1]^{L({\cal O})}\to\R$ une fonction $Z=Z((z_{\ell})_{\ell\in L({\cal O})})$ d\'ependant d'un ensemble de param\`etres $z_{\ell}\in[0,1]$ plac\'es sur les liens $\ell=
(o_{\ell},o'_{\ell})$ reliant deux \`a deux 
les objets d'un ensemble
(abstrait) fini ${\cal O}$.  Alors:

\begin{enumerate}
\item (formule BKAR1)

\BEQ Z_{\cal O}({\bf 1})=\sum_{\F\in {\cal F}({\cal O})} \left( \prod_{\ell\in L(\F)} \int_0^1 dw_{\ell}\right) \left(\left( \prod_{\ell\in L(\F)} \frac{\partial}{\partial z_{\ell}}\right)Z\right)
(\vec{z}(\vec{w})), \label{eq:BKAR1} \EEQ
o\`u: ${\cal F}({\cal O})$ est l'ensemble des for\^ets reliant les objets de $\cal O$; $z_{\ell}(\vec{w}), \ \ell\in L({\cal O})$ est le minimum des param\`etres $w_{\ell'}$ pour $\ell'$
parcourant l'unique chemin de $o_{\ell}$ vers $o'_{\ell}$ si $o_{\ell}$ et $o'_{\ell}$ sont connect\'es par les liens de $\F$, et $z_{\ell}(\vec{w})=0$ sinon. 

\item (formule BKAR2)

On suppose ${\cal O}={\cal O}_1\amalg {\cal O}_2$.  Soit ${\cal F}_{res}({\cal O})$ l'ensemble des for\^ets $\F$ {\em restreintes$^*$} sur $\cal O$, c'est-\`a-dire des for\^ets dont
chaque composante connexe est (i) soit un arbre d'objets de type $1$, appel\'e {\em arbre non enracin\'e$^*$}; soit (ii) un {\em arbre enracin\'e$^*$} contenant un seul sommet de type 2, consid\'er\'e
comme sa racine. Alors
\BEQ Z_{\cal O}({\bf 1})=\sum_{\F\in {\cal F}_{res}({\cal O})} \left( \prod_{\ell\in L(\F)} \int_0^1 dw_{\ell}\right) \left(\left( \prod_{\ell\in L(\F)} \frac{\partial}{\partial z_{\ell}}\right)Z\right)
(\vec{z}(\vec{w})), \label{eq:BKAR2} \EEQ
o\`u  $z_{\ell}(\vec{w}), \ \ell\in L({\cal O})$ est le minimum des param\`etres $w_{\ell'}$ pour $\ell'$
parcourant l'unique chemin dans $\bar{\F}$ de $o_{\ell}$ vers $o'_{\ell}$, $\bar{\F}$ \'etant la for\^et obtenue \`a partir de $\F$ en fusionnant$^*$ toutes les racines de $\F$ en un seul sommet.
\end{enumerate}
\end{Definition}

En pratique $Z_{\cal O}:=Z_{\cal O}({\bf 1})=Z_{\cal O}(1,\ldots,1)$ est une fonctionnelle donn\'ee au d\'epart (d\'ependant de l'ensemble $\cal O$ ainsi qu'implicitement de param\`etres ext\'erieurs comme des constantes de couplage, des champs ext\'erieurs...), d'une
grande complexit\'e combinatoire, c'est-\`a-dire (sans pr\'etendre \`a une d\'efinition pr\'ecise) ne pouvant s'obtenir \`a partir
de fonctionnelles sur  des sous-ensembles de $\cal O$. Dans les deux cas que nous allons
voir, il existe une fa\c con naturelle d'associer \`a $Z_{\cal O}$ une fonctionnelle $Z_{\cal O}((z_{\ell})_{\ell\in L({\cal O})})$ de sorte que $Z_{\cal O}(\vec{z})$ se factorise
sur les composantes connexes du graphe obtenu en supprimant les ar\^etes $\ell$ telles que $z_{\ell}=0$. Plus pr\'ecis\'ement,  la contribution d'une for\^et $\F$ de composantes connexes $\T_1,\ldots,\T_I$
au membre de droite de (\ref{eq:BKAR1}) se r\'e\'ecrit comme le produit \BEQ \prod_{i=1}^I \left( \prod_{\ell\in L(\T_i)} \int_0^1 dw_{\ell}\right) \left(\left( \prod_{\ell\in L(\T_i)} \frac{\partial}{\partial z_{\ell}}\right)Z_{\T_i}\right)
(z_{\ell}((w_{\ell'})_{\ell'\in L(\T_i)}),\ell\in L(\T_i)).\EEQ Le d\'eveloppement restreint (\ref{eq:BKAR2}) \'evite de tester (ou affaiblir) les liens entre deux objets de type 2 lorsque
cela est inutile ou dangereux (cf. application au d\'eveloppement de Mayer
ci-dessous).

\bigskip
%%%%%%%%%%%%

{\underline{Premi\`ere application: d\'eveloppement en clusters horizontal}}

\medskip

Soit $\D^{j}$ l'ensemble des cubes $\Del^j$ d'une \'echelle $j$ donn\'ee. En dimension $D=1$, il s'agit simplement des intervalles
$[iM^{-j},(i+1)M^{-j}]$, $i\in\Z$. En dimension 2, les carr\'es $\Del^j$ s'obtiennent comme faces du r\'eseau  carr\'e usuel sur $\Z^2$, contract\'e ou dilat\'e d'un facteur $M^{-j}$. De mani\`ere g\'en\'erale, $\Del^j=\{x\in\R^D\ ;\ 
n_i\le M^j x_i\le n_i+1\}$ pour certains entiers $n_1,\ldots,n_D$. Les diff\'erents $\Del^j$ donnent une partition de l'espace $\R^D$. Les liens $\ell$ relient deux cubes not\'es
$\Del_{\ell}$ et $\Del'_{\ell}$. Si $x\in\R^D$, on note $\Del^j_x$ le cube d'\'echelle $j$ contenant $x$.

Consid\'erons la fonction de partition d'un mod\`ele de th\'eorie des champs {\em restreinte} \`a une \'echelle $j$ donn\'ee,
\BEQ Z^j:=\int e^{-{\cal L}_{int}(\psi^{j})(x)dx} d\mu(\psi^j).\EEQ

A un choix de param\`etres d'affaiblissement$^*$ $z_{\ell}$ donn\'e, not\'es dans ce cas particulier $s_{\ell}$ ou $s_{\Del_{\ell},\Del'_{\ell}}$, correspond un noyau de covariance
affaibli $C_{\vec{s}}(x,x'):=s_{\Del^j_x,\Del^j_{x'}} C(x,x')$.

Une g\'en\'eralisation en dimension infinie de la formule suivante pour un vecteur gaussien $(X,Y)$, obtenue par transformation de Fourier 
 \BEA \frac{\partial}{\partial s} \langle f(X,Y)\rangle_s &=&\frac{\partial}{\partial s} \int \int f(x,y) \exp -\half \langle C_s^{-1} \left(\begin{array}{c}  x \\  y\end{array}\right),
\left( \begin{array}{c} x\\ y\end{array}\right) \rangle\ dx\ dy \nonumber\\
 &=& \frac{\partial}{\partial s} \int \int \hat{f}(x,y) \exp -\half\langle C_s \eta,\eta\rangle \ d\eta\nonumber\\
 &=& C_s(X,Y) \langle \partial_X\partial_Y F(X,Y)\rangle_s \EEA
donne la formule suivante:

\BEA && Z^{j}=\sum_{\F^{j}\in {\cal F}^j} \left[ \prod_{\ell\in L(\F^j)} \int_0^1 dw_{\ell} \int_{\Del_{\ell}}dx_{\ell} \int_{\Del'_{\ell}}dx'_{\ell} C_{\vec{s}(\vec{w})}(x_{\ell},x'_{\ell})\right] \nonumber\\
&& \qquad \qquad \int d\mu_{\vec{s}(\vec{w})}(\psi) {\mathrm{Hor}}^j \left(e^{-\int {\cal L}(\psi)(x) dx} \right),\EEA
o\`u Hor$^j:=\prod_{\ell\in L(\F^j)} \left( \frac{\del}{\del\psi^j(x_{\ell})} \frac{\del}{\del\psi^j(x'_{\ell})} \right)$ sera appel\'e {\em op\'erateur de d\'eveloppement horizontal}$^*$.

\bigskip
Regardons maintenant comment cette formule s'applique dans un contexte multi-\'echelles. Le d\'eveloppement \`a l'\'echelle $\rho$ s'applique \`a la fonction $\int e^{-\int_V{\cal L}_{int}(\psi^{\to\rho})(x)dx} d\mu(\psi^{\rho})$, dans laquelle les champs de bas
moment $\psi^{\to(\rho-1)}$ sont consid\'er\'es comme des sources. Le produit de ce d\'eveloppement est une somme de  produit de termes factoris\'es sur chaque arbre $\T^{\rho}$ provenant
de la formule de Brydges-Kennedy-Abdesselam-Rivasseau. Chaque terme factoris\'e est une somme de termes du type $C^{\rho}G^{\rho}e^{-\int_{\T^{\rho}} {\cal L}_{int}(\psi^{\to\rho})(x) dx}$, o\`u: $C^{\rho}$ est un produit de propagateurs $C^{\rho}(x_{\ell},x'_{\ell})$ d'\'echelle $\rho$; $G^{\rho}$ est un produit (fini) de champs $\psi^{\to\rho}$. 

En r\'ealit\'e, ce d\'eveloppement s'applique non pas \`a la fonction de partition $Z^{\to\rho}$, mais \`a la fonction $Z^{\to\rho}(\vec{t})$, o\`u les coefficients $t^j_{\Del^j},j\in\Z,
\Del^j\in \D^j$ multipliant diversement les diff\'erentes \'echelles des champs sont introduits en vue du d\'eveloppement multi-\'echelles (cf. \S 2.2). La d\'ecomposition pr\'ec\'edente reste n\'eanmoins valable,
on obtient 
des termes du type $C^{\rho}G^{\rho}e^{-\int_{\T^{\rho}} {\cal L}_{int}(\psi^{\to\rho};\vec{t})(x)dx}$ o\`u la d\'ependance en les param\`etres $t$ est cach\'ee dans le mon\^ome $G^{\rho}$.  

\medskip

Les d\'eveloppements aux diff\'erentes \'echelles s'obtiennent de la m\^eme mani\`ere. D'un point de vue logique, apr\`es avoir compl\'et\'e les d\'eveloppements en clusters aux \'echelles
$\rho,\ldots,j+1$ et obtenu des polym\`eres, il suffirait d'affaiblir par des param\`etres $s$ les liens entre cubes d'\'echelle $j$ situ\'es \`a la base de polym\`eres {\em distincts}. En pratique
cette proc\'edure n'est pas d\'enu\'ee d'ambigu\"it\'e, et plut\^ot que de d\'efinir une proc\'edure pr\'ecise relativement arbitraire, on peut  mettre sans dommage des param\`etres d'affaiblissement $s$ entre tous les intervalles d'\'echelle $j$.

\bigskip

%%%%%%%%%%%%%%%%%%%%%%%%%%%%%%%%%%%%%\`u

{\underline{Deuxi\`eme application: d\'eveloppement de Mayer}}

%%%%%%%%%%%%%%%%%%%%%%%%%%%%%%%%%%%\`u

\medskip

Ce d\'eveloppement permet de s\'eparer la contribution des polym\`eres du vide$^*$ \`a l'\'energie libre, comme nous l'expliquons ci-dessous, et de resommer les parties
locales des polym\`eres divergents en vue de la renormalisation. Contentons-nous ici de montrer comment
effectuer ce d\'eveloppement \`a l'aide de la formule de  Brydges-Kennedy-Abdesselam-Rivasseau. Un {\em polym\`ere} $\P^{j\to}$ multi-\'echelle d'\'echelle minimale $j$ est un ensemble connexe de cubes d'\'echelles $k=j,j+1,\ldots,\rho$
reli\'es par des liens horizontaux (entre cubes de m\^eme \'echelle) ou verticaux, encore dits {\em liens d'inclusion$^*$}, reliant $\Del^k$, $k>j$ \`a l'unique cube $\Del^{k-1}$ d'\'echelle
$k-1$ contenant $\Del^k$. Deux polym\`eres $\P_1^{j\to},\P_2^{j\to}$ sont dits {\em non-$j$-overlapping} (sans recouvrement \`a l'\'echelle $j$ ?) si $\P_1^{j\to}$ et $\P_2^{j\to}$ n'ont pas
de cube en commun \`a l'\'echelle $j$, autrement dit, si $(\P_1^{j\to}\cap\D^j)\cap(\P_2^{j\to}\cap\D^j)
=\emptyset$.  Les objets sont cette fois-ci des polym\`eres, et les liens, des liens de non-overlap qu'on souhaite supprimer. La formule de BKAR permet de ne garder des liens
de non-overlap qu'entre les polym\`eres appartenant \`a un m\^eme arbre de polym\`eres. Le d\'eveloppement fait appara\^itre de mani\`ere g\'en\'erale des {\em for\^ets de polym\`eres}  --
une sorte de superstructure arborescente si l'on pense que les polym\`eres sont eux-m\^emes des arbres --. Les param\`etres d'affaiblissement sont not\'es ici $S$ (pour ne pas les confondre
avec les param\`etres $s$ du d\'eveloppement en clusters horizontal). Le lecteur v\'erifiera facilement que mettre des param\`etres $S$ \`a $0$ \'equivaut \`a autoriser deux polym\`eres \`a se chevaucher
librement, alors que d\'eriver par rapport \`a un param\`etre $S$ implique que les polym\`eres ont au moins un cube en commun. Dans le deuxi\`eme cas, le cube en commun attache les deux polym\`eres
l'un \`a l'autre, supprimant la libert\'e  de d\'eplacement horizontal de l'un par rapport \`a l'autre; l'arbre de polym\`eres ainsi cr\'e\'e se borne alors comme un polym\`ere simple puisqu'il a \'et\'e
rendu connexe (cf. \S 3.2). Dans le premier cas, les contributions des deux polym\`eres se multiplient, permettant la resommation en exponentielle. Si ces polym\`eres poss\`edent des champs
externes, il faut n\'eanmoins les consid\'erer comme des "polym\`eres color\'es", de couleurs diff\'erentes (puisqu'ils ont \'et\'e rendus totalement ind\'ependants l'un par rapport \`a l'autre). It\'erant
le d\'eveloppement \`a chaque \'echelle, on voit qu'il faut
donc travailler de mani\`ere g\'en\'erale avec des "champs color\'es", $\tilde{\psi}^j:\R^D\times \{$couleurs$\}\to\C$. Les corrections aux bornes gaussiennes du \S 3.2 sont mineures (cf.
\cite{MagUnt2} pour les d\'etails).

\begin{Lemme}
Soit ${\mathrm{NonOverlap}}^j(\P_1,\ldots,\P_N):=\prod_{(\P_n,\P_{n'})} {\bf 1}_{\P_n,\P_n' {\mathrm{non-j-overlapping}}}$ et
\BEA && {\mathrm{NonOverlap}}^j(\P_1,\ldots,\P_N;\vec{S}):= \nonumber\\
&& \quad\prod_{(\P_n,\P_{n'})}  \left( (1-S_{\P_n,\P_{n'}})+S_{\P_n,\P_{n'}} {\bf 1}_{\P_n,\P_n' {\mathrm{non-j-overlapping}}} \right). \nonumber\\ \EEA
 Alors
\BEA && {\mathrm{NonOverlap}}^j(\P_1,\ldots,\P_N)= \nonumber\\ && \qquad \sum_{\F\in {\cal F}(\{\P_1,\ldots,\P_N\})} \left( \prod_{\ell\in L({\cal F})} \int_0^1 dW_{\ell}\right)
 \left(\prod_{\ell\in L({\cal F})} \frac{\partial}{\partial S_{\ell}}\right) {\mathrm{NonOverlap}}^j(\vec{S}(\vec{W})), \nonumber\\ \EEA
o\`u ${\cal F}(\{\P_1,\ldots,\P_N\})$ est l'ensemble des for\^ets reliant les polym\`eres $\P_1,\ldots,\P_N$.
\end{Lemme}

En pratique on distinguera entre les polym\`eres de type 1 (poss\'edant $<N_{ext,max}$ champs externes) et ceux de type 2 (poss\'edant $\ge N_{ext,max}$ champs externes) -- puisque le d\'eveloppement
de Mayer sert en principe \`a resommer les parties locales des polym\`eres divergents --, et  on n'affaiblira
la condition de non-overlap qu'entre les intervalles dans lesquels aucun champ externe n'est situ\'e, ceci afin d'\'eviter l'accumulation de champs externes dans le m\^eme intervalle que pourraient
cr\'eer
les $S$-d\'erivations autrement, et on appliquera la 2\`eme formule de Brydges-Kennedy-Abdesselam-Rivasseau au lieu de la 1\`ere. Nous n'\'ecrirons pas la formule explicitement.

%%%%%%%%%%%%%%%%%%%%%%%%%%%%

\subsection{D\'eveloppement en clusters vertical}

%%%%%%%%%%%%%%%%%%%%%%%%%%

Ce d\'eveloppement est en r\'ealit\'e un d\'eveloppement de Taylor partiel dans chaque cube d'une \'echelle donn\'ee. En ce sens ce n'est pas r\'eellement un d\'eveloppement en cluster, et la formule de Brydges-Kennedy-Abdesselam-Rivasseau ne s'applique pas. N\'eanmoins, comme nous allons voir, il conduit \`a une s\'eparation effective de l'ensemble des cubes $\amalg_j \D^j$ en morceaux
disjoints appel\'es {\em polym\`eres}, qu'on peut voir comme des "clusters" ou amas multi-\'echelles, et les param\`etres $t\in[0,1]$ ci-dessous jouent un r\^ole tr\`es similaire aux param\`etres
d'affaiblissement $s$ ou $S$ des paragraphes pr\'ec\'edents. La terminologie n'est pas fix\'ee, on parle soit de d\'eveloppement en cluster
vertical$^*$, soit (plus justement peut-\^etre) de "momentum-decoupling expansion" ou "d\'eveloppement sur le couplage entre \'echelles".

\begin{Definition}[param\`etres verticaux]

On note $\vec{t}$ un ensemble de param\`etres $t_{\Del^j}^j$, $j\in\Z$, $\Del^j\in\D^j$, associ\'e chacun \`a un cube donn\'e. De mani\`ere \'equivalente, $x\mapsto t_x^j:=t_{\Del^j_x}^j$, o\`u
$\Del^j_x$ est l'unique cube d'\'echelle $j$ contenant un point $x$, d\'efinit une fonction localement constante dans chaque cube.

\end{Definition}

Voici une premi\`ere d\'efinition possible.

\begin{Definition}[lagrangien habill\'e$^*$ ]

Soit ${\cal L}_{int}(\psi)(x)=\prod_{i=1}^I \psi_i(x)$ un mon\^ome contenu dans le lagrangien d'interaction, coupl\'e \`a une constante de couplage $\lambda$, et ${\cal L}_{int}^{\to\rho}(\psi)(x):=\prod_{i=1}^I \psi_i^{\to\rho}(x)$
sa troncature$^*$ ultra-violette. Alors le lagrangien habill\'e$^*$ associ\'e est
\BEQ {\cal L}_{int}^{\to\rho}(\psi;\vec{t})(x):=\lambda^{\rho} \prod_{i=1}^I (T\psi_i)^{\to\rho}(x)+\sum_{\rho'\le\rho} \lambda^{\rho'-1} (1-(t_x^{\rho'})^I)\prod_{i=1}^I (T\psi_i)^{\to(\rho'-1)}(x). \label{eq:2.10} \EEQ

\end{Definition}

Lorsque la constante de couplage $\lambda$ n'est pas renormalis\'ee, la formule se simplifie, la constante $\lambda$ \'etant en facteur \`a la fois dans ${\cal L}_{int}(\psi)$ et
dans ${\cal L}_{int}(\psi;\vec{t})$; comme pour les d\'eveloppements en cluster, on retrouve alors la th\'eorie initiale lorsque $\vec{t}=1$, autrement dit,  ${\cal L}_{int}(\psi)(x)=
{\cal L}_{int}(\psi;{\bf 1})(x)$. Mais en g\'en\'eral, la renormalisation induit un flot du param\`etre $\lambda$,
$\lambda^j$ \'etant la constante de couplage effective relative \`a l'\'echelle $j$. Cette constante effective
s'obtient en pratique en sommant les contre-termes de toutes les \'echelles $\ge j$. Autrement dit, le {\em contre-terme
d'\'echelle $j$}, $-(\lambda^{j-1}-\lambda^{j})$, compense les divergences de la th\'eorie dues aux sous-diagrammes
dont l'\'echelle la plus basse est pr\'ecis\'ement \'egale \`a $j$ \footnote{Plus pr\'ecis\'ement, $\lambda^{j-1}-\lambda^j$
est \'egale \`a la somme des contributions des {\em parties locales} des {\em polym\`eres} dont l'\'echelle la plus basse
est \'egale \`a $j$ (cf. infra).}.

La formule (\ref{eq:2.10}) se r\'e\'ecrit \`a l'aide des contre-termes d'\'echelle $j$, en faisant une resommation
d'Abel,
\BEQ  {\cal L}_{int}^{\to\rho}(\psi;\vec{t})(x)=\sum_{\rho'\le \rho+1} (\lambda^{\rho'-1}-\lambda^{\rho'})
\sum_{\rho''\le \rho'} (1-(t_x^{\rho'' })^I) \prod_{i=1}^I (T\psi_i)^{\to(\rho''-1)}(x),\EEQ
o\`u l'on a pos\'e $\lambda^{\rho+1}=0.$

\medskip

Dans le cas des chemins rugueux renormalis\'es, la mise en ordre normal de Fourier fait appara\^itre un lagrangien d'interaction d\'ecompos\'e en \'echelles. Pour un lagrangien du
type ${\cal L}_{int}(\psi)(x)=\sum_{(j_i)_{i\in I}} K^{(j_i)} \prod_{i=1}^I \psi_i^{j_i}(x)$, la formule d'habillage se g\'en\'eralise ainsi:

\BEA && {\cal L}_{int}^{\to\rho}(\psi;\vec{t})(x):=\lambda^{\rho} \sum_{0\le (j_i)_{i\le I}\le \rho} K^{(j_i)} \prod_{i=1}^I (T^{\to\rho}\psi_i)^{j_i}(x) \nonumber\\
&&\qquad +\sum_{\rho'\le\rho} \lambda^{\rho'-1} (1-(t_x^{\rho'})^I) \sum_{0\le (j_i)_{i\le I}\le \rho'-1} K^{(j_i)} \prod_{i=1}^I (T^{\to(\rho'-1)}\psi_i)^{j_i}(x). \nonumber\\ \EEA

\bigskip

Nous pouvons maintenant introduire le d\'eveloppement vertical. De la m\^eme mani\`ere que le d\'eveloppement horizontal s'obtient par un d\'eveloppement de Taylor
habile en les variables $s_{\Del_{\ell},\Del'_{\ell}}$ autour de la valeur $s=0$, le d\'eveloppement vertical s'obtient par un simple d\'eveloppement de Taylor \`a l'ordre $N_{ext,max}$
en les variables $t_{\Del}$. On peut formaliser ainsi ce d\'eveloppement :

\begin{Definition}[op\'erateurs de d\'eveloppement vertical]
Soit $j\in\Z$ et $\T^j$ un arbre reliant des cubes d'\'echelle $j$; on note $\Vert^j_{\T_j}$ l'op\'eration de d\'eveloppement de Taylor \`a l'ordre $N_{ext,max}$ dans chacun des cubes de $\T_j$,
\BEA && \Vert^j \left[ f(t^j_{\Del^j},\Del^j\in\T^j)\right]=\prod_{\Del^j\in\T^j} \left\{ {\bf 1}\big|_{t^j_{\Del^j}=0}+\partial_{t^j_{\Del^j}}\big|_{t^j_{\Del^j}=0}+\ldots+
 \right.\nonumber\\
 && \left. \quad 
\partial^{N_{ext,max}-1}_{t^j_{\Del^j}}\big|_{t^j_{\Del^j}=0}+\int_0^1 dt^j_{\Del^j} \frac{(1-t^j_{\Del^j})^{N_{ext,max}-1}}{(N_{ext,max}-1)!} \partial_{t^j_{\Del^j}}^{N_{ext,max}}\right\}
f(t^j_{\Del^j},\Del^j\in\T^j). \nonumber\\ \EEA
\end{Definition}

La formule de Taylor avec reste int\'egral donne  $f({\bf 1})=\Vert^j f$. Appliqu\'ee \`a l'\'echelle $\rho$ \`a la trace sur un arbre $\T^{\rho}$   du d\'eveloppement en cluster horizontal \`a l'\'echelle $\rho$, elle produit une
somme de produits de termes du type $G((t^{\rho}_{\Del^{\rho}})_{\Del^{\rho}\in\T^{\rho}})e^{-\int_{\T^{\rho}}{\cal L}_{int}(\psi;(t^{\rho}_{\Del^{\rho}})_{\Del^{\rho}\in\T^{\rho}})(x) dx}$,
multipli\'es par un produit de propagateurs d'\'echelle $\rho$. S\'eparons  les champs $(T\psi)^{\to\rho}(x)$ apparaissant dans $G$ en $\psi^{\rho}(x)+t^{\rho}_x (T\psi)^{\to(\rho-1)}(x)$,
et voyons le r\'esultat sur le support de l'arbre $\T^{\rho}$ de l'action des diff\'erents termes dans les op\'erateurs de d\'eveloppement vertical $\Vert^{\rho}$. Le principe g\'en\'eral est que chaque
d\'erivation $\partial_{t^{\rho}_{\Del^{\rho}}}$ produit un champ de bas moment $(T\psi)^{\to(\rho-1)}$. On trouve les diff\'erents cas suivants:

\begin{enumerate}
\item[(i)] Choisissons le terme de degr\'e $0$, ${\bf 1}\big|_{t^{\rho}_{\Del^{\rho}}=0}$, dans chaque cube de $\T^{\rho}$. Alors tous les termes $t^{\rho}_x(T\psi)^{\to(\rho-1)}(x)$ dans $G$
ont \'et\'e mis \`a z\'ero. L'expression r\'esultante ne contient donc pas de sources ou champs externes de moment $\le \rho-1$, c'est une constante qu'on appelle {\em polym\`ere du vide$^*$}. En d\'eveloppant en s\'erie $e^{-\int_{\T^{\rho}}{\cal L}_{int}(\psi^{\rho})(x) dx}$ l'exponentielle dans laquelle tous les champs de bas moment ont \'et\'e mis \`a z\'ero, on retrouverait 
la somme formelle des {\em diagrammes du vide$^*$} d'\'echelle $\rho$ de la th\'eorie perturbative.   Mais en th\'eorie perturbative, les diagrammes du vide ne sont pas consid\'er\'es; en effet, leur resommation
conduit \`a une exponentielle qui   change simplement le facteur de normalisation de la mesure, et
ne modifie donc pas les fonctions \`a $n$ points. Les contraintes de non-overlap entre polym\`eres ne permettent pas une resommation exponentielle directe en th\'eorie
constructive, mais  le d\'eveloppement de Mayer permet de lever ces contraintes de non-overlap, faisant appara\^\i tre un facteur exponentiel \'egal \`a $e^{|V|M^{\rho} f^{\rho}_V}$, o\`u
$f^{\rho}_V$ converge quand $|V|\to\infty$ vers l'\'energie libre par degr\'e de libert\'e d'\'echelle $\rho$, not\'ee $f^{\rho}$, et d'ordre $O(\lambda)$. 

{\em Exemple.} 

\item[(ii)] A l'autre extr\^eme, supposons que le reste int\'egral ait \'et\'e choisi pour au moins l'{\em un} des cubes dans $\T^{\rho}$. Le polym\`ere poss\`ede donc au moins $N_{ext,max}$
champs externes. Le comptage de puissance expliqu\'e en section 1 montre que  les diagrammes de Feyman poss\'edant suffisamment de pattes externes sont superficiellement convergents.
On d\'efinit alors $N_{ext,max}-1$ comme le nombre maximum de pattes externes d'un diagramme de Feynman superficiellement divergent. 

\item[(iii)] Les cas interm\'ediaires se rangent plut\^ot dans la premi\`ere ou deuxi\`eme cat\'egorie, suivant  le nombre de champs externes. Si celui-ci est   $< N_{ext,max}$,  on
retire \`a ce polym\`ere sa {\em partie locale} -- ou {\em \'evaluation  \`a moments externes nuls}. Graphiquement, l'\'evaluation \`a moments externes nuls est \'equivalente \`a {\em d\'eplacer
tous les champs externes au m\^eme point} \cite{FVT}. Apr\`es le d\'eveloppement de Mayer, cette extraction de parties locales s'av\`ere \'equivalente \`a une
renormalisation des param\`etres. Au contraire, s'il est $\ge N_{ext,max}$, on ne fait rien; le polym\`ere est consid\'er\'e comme superficiellement convergent.

\end{enumerate}

%%%%%%%%%%%%%%%%%%%%%%%%%%%%\`u

\subsection{D\'eveloppement en clusters multi-\'echelles}

%%%%%%%%%%%%%%%%%%%%%%%%%%%%%

En th\'eorie perturbative des champs, un comptage de puissance "na\"if" (mono-\'echelle) permet de rep\'erer les structures externes possibles des graphes superficiellement divergents. Le d\'ebut
de leur d\'eveloppement de Taylor \`a moments externes nuls (appel\'e {\em partie locale} dans le language de la th\'eorie constructive) est mis \`a part et  resomm\'e. Leur contribution s'obtient de mani\`ere
\'equivalente en rajoutant un contre-terme au lagrangien, ou encore en rempla\c cant les param\`etres nus$^*$ par les param\`etres renormalis\'es.

L'id\'ee est exactement la m\^eme en th\'eorie constructive, \`a ceci pr\`es que: les diagrammes de Feynman sont remplac\'es par des polym\`eres multi-\'etages; la resommation des parties locales ne
peut se faire qu'{\em apr\`es} avoir lev\'e la contrainte de non-overlap entre les polym\`eres (d\'eveloppement de Mayer). La proc\'edure pr\'ecise est facile \`a comprendre, mais les notations
pr\'ecises sont  lourdes, et le d\'eveloppement de Mayer complique de mani\`ere inessentielle les bornes constructives de la section suivante puisqu'il impose de sommer sur des {\em arbres
de polym\`eres} au lieu de polym\`eres.  Nous nous contenterons donc de pr\'esenter ici les arguments essentiels, en all\'egeant les notations. Le d\'eveloppement se fait par r\'ecurrence, en 
partant de l'\'echelle la plus haute, $\rho$, suivant le sch\'ema: $\Hor^{\rho}\longrightarrow \Vert^{\rho}\rightsquigarrow \Hor^{\rho-1}\longrightarrow \Vert^{\rho-1}\rightsquigarrow\ldots,$
o\`u $\Hor^j$, resp. $\Vert^j$, symbolise le d\'eveloppement en clusters horizontal, resp. vertical \`a  l'\'echelle $j$.
Chaque fl\`eche interm\'ediaire $\rightsquigarrow$ recouvre en fait trois op\'erations \'el\'ementaires successives. Pla\c cons-nous \`a l'\'echelle $j$, c'est-\`a-dire juste apr\`es le d\'eveloppement vertical
d'\'echelle $j$. On suppose (hypoth\`ese de r\'ecurrence) que la fonction de partition a \'et\'e r\'e\'ecrite comme un produit $\prod_{k=j+1}^{\rho} e^{|V|M^k f^{k\to\rho}(\lambda)}$
(contribution \`a la fonction de partition des degr\'es de libert\'e d'\'echelles $>j$), multipli\'e par  une somme sur des polym\`eres (en fait: sur des arbres de polym\`eres) d'\'echelle la plus basse $\ge j$ sans overlap \`a l'\'echelle $j$, i.e.
\BEQ \sum_{N=1}^{\infty} \frac{1}{N!} \sum_{{\mathrm{non-j-overlapping}}\ \P_1,\ldots,\P_n\in {\cal P}^{j\to}} \prod_{n=1}^N F^j_{HV}(\P_n;\psi), \label{eq:avantMayer} \EEQ obtenue apr\`es les d\'eveloppements
horizontal (H) et vertical (V) \`a l'\'echelle $j$. 

\begin{enumerate}
\item Si le polym\`ere $\P_n$ poss\`ede $<N_{ext,max}$ champs externes, on met de c\^ot\'e sa partie locale, obtenue en d\'epla\c cant tous les champs externes au m\^eme point. Ce qui reste est 
consid\'er\'e  artificiellement (pour la coh\'erence du sch\'ema) comme un polym\`ere avec $\ge N_{ext,max}$ champs externes puisqu'il est convergent.

\item On applique le d\'eveloppement de Mayer \`a l'\'echelle $j$ \`a l'expression (\ref{eq:avantMayer}).
\item On resomme les polym\`eres du vide en une exponentielle $e^{|V|M^j f^{j\to\rho}(\lambda)}$, et les parties locales des polym\`eres divergents (avec un nombre de champs externes
compris entre $1$ et $N_{ext,max}-1$ donc) en un contre-terme d'\'echelle $j$ qu'on remet dans le lagrangien.

\end{enumerate}

%%%%%%%%%%%%%%%%%%%%%%%%%%

\section{Bornes constructives}

%%%%%%%%%%%%%%%%%%%%

On retrouve ici  le comptage de puissance de la section 1, mais avec l'obligation de savoir sommer sur tous les "graphes" simultan\'ement, qui complique singuli\`erement
le probl\`eme. Le principe g\'en\'eral
est que le d\'eveloppement multi-\'echelles a produit une somme de termes du type\\ $\prod_{j\in\Z} \int d\mu_{\vec{s}^j}(\psi^j) C^j G^j e^{-\int {\cal L}_{int}(\psi;\vec{t})(x)dx}$,
abr\'eg\'es en $C^j G^j e^{-\int {\cal L}_{int}}$, o\`u:
$\vec{s}^j,t^j_{\Del^j}$ sont les param\`etres des d\'eveloppements en cluster horizontaux et verticaux; $C^j$ est un produit de propagateurs d'\'echelle $j$; $G^j$ est un produit de champs
d'\'echelle $j$ produits par des d\'eveloppements en cluster d'\'echelle arbitraire; et ${\cal L}_{int}(\psi;\vec{t})$ est un lagrangien multi-\'echelle avec des
param\`etres renormalis\'es qu'on d\'etermine de mani\`ere r\'ecursive. Il faut donc commencer par estimer ces param\`etres renormalis\'es, qu'on obtient comme solution d'\'equations implicites.
On peut alors calculer l'\'energie libre ou les fonctions \`a $n$ points. Tous ces calculs reposent sur les m\^emes principes: (i) on borne l'exponentielle $e^{-\int {\cal L}_{int}(\psi;\vec{t})(x) dx}$, si possible (mais pas toujours) par $1$ (ou en tout cas une constante par cube), et on se ram\`ene par l'in\'egalit\'e de Cauchy-Schwarz \`a calculer $\prod_{j\in\Z} \int d\mu_{\vec{s}^j}(\psi^j) (C^j G^j)^2$;
(ii) on utilise la formule de Wick en consid\'erant toutes les contractions possibles, ce qui produit des sommes de diagrammes de Feynman multi-\'echelles (avec des contraintes de
non-overlap); (iii) on somme sur toutes les for\^ets de cluster (et for\^ets de polym\`eres de Mayer) possibles ({\em bornes gaussiennes$^*$}), en tenant compte des {\em facteurs
combinatoires} de la formule de Leibniz donnant la d\'eriv\'ee d'un produit (chaque d\'erivation $\frac{\partial}{\partial s^j_{\Del^j,(\Del^j)'}}$ agissant comme $\frac{\del}{\del\psi(x_{\ell})}\frac{\del}{\del
\psi(x'_{\ell})}$, ou $\frac{\partial}{\partial t_j^{\Del^j}}$, agissant sur un produit de champs). Si l'on avait d\'evelopp\'e l'exponentielle en s\'erie, cette somme divergerait
en raison de facteurs exponentiels d\^us \`a l'accumulation de champs d'\'echelle $j$ dans une zone de taille de l'ordre de $M^{-j}$. Il faut v\'erifier que les d\'eveloppements en cluster
\'evitent l'apparition de ces facteurs exponentiels. En pratique, des factorielles apparaissent si l'on n'y prend garde, en raison de l'accumulation de {\em champs moyenn\'es de bas moment} (cf.
explications ci-dessous); ces champs moyenn\'es doivent \^etre {\em domin\'es} directement par une exponentielle d\'ecroissante provenant de l'action $e^{-\int {\cal L}_{int}(\psi;\vec{t})(x)dx}$.
Contrairement aux {\em bornes gaussiennes} et aux facteurs combinatoires, essentiellement universels, la {\em domination} d\'epend fortement du mod\`ele consid\'er\'e, en particulier des
dimensions d'\'echelle des champs; elle n'est possible que sous des hypoth\`eses de {\em stricte positivit\'e du lagrangien}, proches dans l'esprit d'hypoth\`eses de convexit\'e. Bien entendu, toutes les bornes d\'ependent de mani\`ere essentielle du flot du groupe de renormalisation.

Tous ces probl\`emes combin\'es contribuent \`a rendre les bornes constructives illisibles. Malgr\'e de nombreux efforts de simplification, de syst\'ematisation ou de r\'e\'ecriture alg\'ebrique \cite{Abd,GuMaRi,MagRiv},
il semble qu'un principe de non-r\'eduction des difficult\'es op\`ere ici. N\'eanmoins, dans le cadre de ce "mode d'emploi", on peut pr\'esenter les  arguments essentiels revenant de mani\`ere
r\'ecurrente dans tous les mod\`eles bosoniques consid\'er\'es, en esp\'erant que les lignes qui suivent permettront au lecteur de s'orienter dans les articles complets. Nous ne discutons
pas ici la {\em domination}, pr\'ef\'erant l'introduire dans la section suivante sur des mod\`eles concrets.

\subsection{Comptage de puissance constructif}

Rappelons que la covariance de la composante d'\'echelle $j$ d'un champ de dimension $\beta$ est born\'ee par
 $|\langle\psi^j(x)\psi^j(y)\rangle|\le C_r\frac{M^{2\beta j}}{(1+M^j|x-y|)^r}$. Si l'on oublie la d\'ecroissance spatiale, chaque composante de champ $\psi^j$ contribue
 un facteur $M^{\beta j}$, explicite dans le comptage de puissance des diagrammes multi-\'echelles (cf. section 1). 
 
Voyons les diff\'erents champs produits par les d\'eveloppements en cluster horizontaux et verticaux: on obtient:

\begin{itemize}
\item[(i)] des {\em propagateurs} $C^j(x,y)=\langle \psi^j(x)\psi^j(y)\rangle$ produits par le d\'eveloppement horizontal, d'ordre de grandeur $M^{2\beta_j}$.
\item[(ii)] des {\em champs de haut moment} produits par les d\'eveloppements horizontaux comme verticaux; les op\'erateurs horizontaux $\frac{\del}{\del \psi^j(x_{\ell})}$ ou verticaux
$\frac{\del}{\del t^j_{\Del^j}}$ appliqu\'es \`a l'exponentielle $e^{-\int {\cal L}_{int}}$ ou \`a un  produit de champs $G$ (provenant des d\'eveloppements en cluster des \'echelles sup\'erieures) 
peuvent faire sortir en particulier des champs  $\psi^k(x),x\in\Del^j$, d'\'echelle $k>j$. On distingue alors l'{\em \'echelle de production$^*$} $j$ de ces champs
de leur {\em \'echelle propre$^*$} $k$. Pour les {\em bornes gaussiennes} on s\'epare en pratique $\psi^k(x)$, $x\in\Del^j$ en somme de {\em champs restreints$^*$} Res$^j_{\Del^k}\psi^k(x):=
{\bf 1}_{x\in\Del^k} \psi^k(x)$, o\`u $\Del^k$ d\'ecrit les $M^{D(k-j)}$ cubes d'\'echelle $k$ contenus dans $\Del^j$. Rappelons que ces champs de haut moment sont un sous-produit de la
m\'ethode constructive; ils n'apparaissent pas quand on \'etudie un diagramme de Feynman multi-\'echelle donn\'e.
\item[(iii)] des champs de bas moment $\psi^h(x),x\in\Del^j$  $(h\le j)$ produits par les m\^emes d\'eveloppements. Comme dans le comptage de puissance des diagrammes de Feynman multi-\'echelle,
on fait {\em comme si} $\psi^h$ \'etait d'\'echelle $j$. Autrement, le facteur $M^{\beta h}$ apport\'e par la composante $\psi^h$ est d\'ecompos\'e en $M^{\beta j}\cdot M^{-\beta(j-h)}$.
Si l'on est tout \`a fait pr\'ecis, ces champs de bas moment ont non pas {\em deux} \'echelles mais {\em trois \'echelles}, en raison du fait qu'ils peuvent \^etre red\'eriv\'es \`a plusieurs
\'echelles successives. Soit $k$ l'{\em \'echelle de production$^*$}, correspondant \`a la premi\`ere d\'erivation $\frac{\del}{\del \psi^k(x_{\ell})}$ ou 
$\frac{\del}{\del t^k_{\Del^k}}$ ayant fait sortir le champ $(T\psi)^{\to(k-1)}(x)=\psi^{(k-1)}(x)+t^{k-1}_x \psi^{(k-2)}(x)+\ldots$. A l'\'echelle suivante $k-1$, la d\'erivation
$\frac{\partial}{\partial t^{k-1}_x}$ peut agir \`a son tour sur $(T\psi)^{\to(k-1)}(x)$, produisant le champ $(T\psi)^{\to(k-2)}$, et ainsi de suite jusqu'\`a une \'echelle $j$, "poussant"
le champ de bas moment initial $(T\psi)^{\to k}$ vers le bas, puisque ses \'echelles les plus hautes sont rabot\'ees au fur et \`a mesure. Par hypoth\`ese le champ $(T\psi)^{\to(j-1)}(x)$
n'est pas touch\'e par les (\'eventuelles) d\'erivations $\frac{\partial}{\partial t^{j-1}_x}$. Il faut bien tenir compte dans les {\em facteurs  combinatoires} de l'\'eventuelle possibilit\'e
qu'il soit  red\'eriv\'e \`a des  \'echelles inf\'erieures,  mais c'est un probl\`eme d'ordre diff\'erent (plus simple peut-\^etre) que nous passerons sous silence. Pour les {\em bornes gaussiennes},
on d\'ecompose le champ $(T\psi)^{\to(j-1)}(x)$ en ses diff\'erentes \'echelles $h<j$. Les composantes r\'esultantes, $\psi^h(x)$, ont donc une {\em \'echelle de production$^*$} $k$, une {\em  \'echelle de derni\`ere
$t$-d\'erivation cons\'ecutive} $j$, et une {\em \'echelle propre} $h$. L'image employ\'ee dans \cite{MagUnt2} est celle d'un ascenseur emmenant le champ $(T\psi)^{\to(k-1)}$ jusqu'\`a l'\'echelle $j$,
puis le laissant tomber en chute libre et \'eclater en ses diff\'erentes composantes $\psi^h$, $h<j$. D'o\`u les termes imag\'es de {\em dropping scale} ({\em \'echelle de l\^achage} ?) au lieu de
l'appellation pr\'ecise mais impronon\c cable d'{\em
   \'echelle de derni\`ere
$t$-d\'erivation cons\'ecutive}, et d'{\em \'echelle de chute libre} ("free falling scale") en concurrence avec \'echelle propre.

Il faut donc \'ecrire en fait $M^{\beta h}=M^{\beta k}\cdot M^{-\beta(k-j)}\cdot M^{-\beta(j-h)}$, les deux {\em facteurs de ressort$^*$}, $M^{-\beta(k-j)}$ et $M^{-\beta(j-h)}$, \'etant utilis\'es
s\'epar\'ement dans deux contextes diff\'erents; on ne s'int\'eressera ici qu'au deuxi\`eme facteur, $M^{-\beta(j-h)}$. En \'eclatant en leurs diff\'erentes composantes, les champs de bas moment
l\^ach\'es \`a l'\'echelle $j$ produisent potentiellement une accumulation de composantes $\psi^h$ (champs d'\'echelle propre $h$) dans un seul et m\^eme cube $\Del^h$; leur nombre maximum
est de l'ordre de $M^{D(j-h)}$, \'egal au nombre de cubes d'\'echelle $j$ contenus dans $\Del^h$. Un calcul un peu sommaire montre les dangers possibles de cette accumulation.
La formule de Wick donne $\esper X^n=1\cdot 3\ldots (n-1)=\frac{n!}{2^{n/2} (n/2)!}\approx \sqrt{n^n}$ ($n$ pair) pour une  variable gaussienne standard $X$, autrement dit
un facteur de l'ordre de $\sqrt{n}$ par variable. Le facteur de d\'ecroissance polyn\^omial \'etant ici de l'ordre de 1 puisque tous les champs $\psi^h$ sont dans le m\^eme cube d'\'echelle $h$,
on a un facteur de l'ordre de $\sqrt{M^{D(j-h)}}$ par champ, multipli\'e par le facteur de ressort $M^{-\beta(j-h)}$, au total un facteur $< 1$ {\em \`a condition que $\beta> D/2$}.
Dans le cas contraire ($\beta\le D/2$), on retire au champ $\psi^h(x)$ sa moyenne sur le cube $\psi^{\Del^j}$ o\`u il a \'et\'e l\^ach\'e. Le r\'esultat, 
\BEQ \del^j\psi^h(x):=\psi^h(x)-\frac{1}{|\Del^j|} \int_{\Del^j}  \psi^h(x)dx \EEQ
appel\'e {\em champ secondaire$^*$}, est d'un ordre inf\'erieur au champ $\psi^h$, au sens o\`u le facteur de ressort n'est plus $M^{-\beta(j-h)}$ mais $M^{-\tilde{\beta}(j-h)}$ avec $\tilde{\beta}=\beta+1$.
Parfois -- comme dans le cas de la th\'eorie $\phi^4$ non massive avec cut-off ultra-violet \cite{FMRS} -- il arrive qu'il faille utiliser une proc\'edure de soustraction de moyenne un peu plus astucieuse {\tt ???}, de sorte que $\tilde{\beta}=\beta+2$. Dans tous les cas, une proc\'edure ou une autre permet d'obtenir un exposant $\tilde{\beta}$ (diff\'erant de $\beta$ par un entier positif)
tel que $\tilde{\beta}>D/2$. On peut poser $\tilde{\beta}=\beta$ si $\beta>D/2$ d\`es le d\'epart (auquel cas le champ secondaire est \'egal au champ initial $\psi^h$), de fa\c con \`a regrouper
tous les cas.  Dans la suite on appellera {\em champ moyenn\'e de bas moment} la diff\'erence $\psi^{\to j}(x)-\del^j\psi^{\to j}(x)$ si $\beta\le D/2$, et on la notera $\psi^{\to j}(\Del^j)$.

\end{itemize}

%%%%%%%%%%%%%%%%%%

\subsection{Bornes gaussiennes}

Ces bornes permettent de traiter les propagateurs, les champs de haut moment et les champs secondaires de bas moment, mais pas les champs moyenn\'es de bas moment, d'accumulation
dangereuse, qui seront trait\'es
s\'epar\'ement dans le paragraphe sur la domination.

Ces bornes sont essentiellement ind\'ependantes du mod\`ele (\`a ceci pr\`es qu'elles d\'ependent bien entendu du flot du groupe de renormalisation qui modifie les constantes). Elles
ne pr\'esentent pas de difficult\'e particuli\`ere. Comme dans la section 6 de \cite{MagUnt2}, on va montrer comment les obtenir par \'etapes en partant de la formule de Wick.

\bigskip

\underline{1. Formule de Wick}

\medskip

Rappelons que, si $X_1,\ldots,X_{2N}$ sont des variables gaussiennes,
 \BEQ \langle X_1\ldots X_{2N}\rangle=\sum_{{\mathrm{pairings}}\ \Pi} X_{\Pi} \label{eq:Wick},\EEQ
  o\`u $\Pi=\{(i_1,i_2),\ldots,(i_{2N-1},i_{2N})\}$ varie dans l'ensemble des "pairings" (appariements ?) des variables $(X_i)$, et $X_{\Pi}=\langle
 X_{i_1}X_{i_2}\rangle\ldots\langle X_{i_{2N-1}i_{2N}}\rangle$. On en d\'eduit facilement, en consid\'erant successivement les diff\'erents "pairings" possibles de $X_1$, puis de $X_2$, etc.:
 \BEQ |\langle X_1\ldots X_{2N}\rangle|\le \prod_{i=1}^{2N-1} \left[ 1+\sum_{j>i} |\langle X_i X_j\rangle |\right] \EEQ
 ou encore par un "rescaling" \'evident des variables $(X_i)$,
 \BEQ |\langle X_1\ldots X_{2N}\rangle|\le K^{-n} \prod_{i=1}^{2N-1} \left[ 1+K\sum_{j>i} |\langle X_i X_j\rangle |\right] \EEQ
pour toute constante $K>0$. Supposant que les covariances $\langle X_i X_j\rangle$ soient toutes du m\^eme ordre de grandeur, cette derni\`ere borne est optimale quand $K$ est choisie
de sorte que les facteurs
$K\langle X_i X_j\rangle$ soient de l'ordre de $1$.

\bigskip

 \underline{2. Bornes mono-\'echelle}
 
\medskip 
 
On fixe une \'echelle $j$ et on consid\`ere uniquement les champs $X=\psi^j(x_{\ell})$ ins\'er\'es dans les propagateurs $C^j$ provenant du d\'eveloppement cluster horizontal \`a l'\'echelle $j$. Na\"ivement
on pourrait penser que le lemme de Wick est inutile puisqu'il suffit de majorer le produit de propagateurs. En fait (en utilisant la factorisation sur les for\^ets) l'on doit sommer sur tous les arbres possibles, ce qui revient
approximativement \`a majorer des expressions du type $\sum_{\Pi} |X_{\Pi}|$, o\`u: $\Pi$ d\'ecrit l'ensemble des "pairings" d'un nombre {\em arbitraire} de champs $\psi^j(x_i),i=1,2,\ldots,2N, x_i\in
\Del^j_i$ (les cubes $\Del^j_i$ n'\'etant pas n\'ecessairement distincts),
avec trois contraintes uniquement: (i) $\Pi$ relie tous les cubes $\Del^j_i$, autrement dit, les liens entre les cubes dans lesquels se situent les champs forment un graphe {\em connexe};
(ii) le nombre $N^j(\Del^j_i)$ de champs $\psi^j(x_i)$ contenus dans un cube fix\'e $\Del^j_i$ est born\'e par $Cn^j(\Del^j_i)$, o\`u $n^j(\Del^j_i)$ est le degr\'e de connectivit\'e du
cube $\Del^j_i$, autrement dit, un plus le nombre de cubes connect\'es \`a $\Del^j_i$; (iii) un certain cube fix\'e $\Del_0^j$ appartient au graphe de cubes. 
 
Ces contraintes sont faciles \`a comprendre: (i) provient de la factorisation de la fonction de partition sur les arbres (ou composantes connexes) du d\'eveloppement en cluster horizontal; (ii)
les champs proviennent des op\'erateurs de d\'erivation Hor$^j$ et $\Vert^j$, il y a en a au plus $In^j(\Del^j_i)$, resp. $IN_{ext,max}$ par cube, o\`u $I$ est le degr\'e de l'interaction ${\cal L}_{int}$;
(iii) s'obtient en fixant un cube, op\'eration n\'ecessaire en raison de l'invariance globale par translation (en d'autres termes, de l'extensivit\'e de l'\'energie libre).   
 
On obtient alors (cf. \cite{MagUnt2}, \'eq. (6.5)), en choisissant un arbre couvrant$^*$ le graphe de cubes et en explorant les sommets un \`a un en partant du cube fix\'e $\Del_0$:

\BEQ \sum_{\Pi} \prod_{\Del} (1+N^j(\Del))^{-1} |X_{\Pi}| \le \left( 1+\sup_{\Del\in\D^j}  \sum_{\Del'\in\D^j} \sup_{x\in\Del,x'\in\Del'} \langle \psi^j(x)\psi^j(x')
\rangle \right)^{3N}.\EEQ

L'insertion du facteur $(1+N^j(\Del))^{-1}$ permet de sommer sur tous les champs contenus dans un cube $\Del$ donn\'e. A priori il sort du chapeau, et sans ce facteur, on obtient
ce qu'on a appel\'e historiquement des {\em factorielles locales$^*$} du type $N^j(\Del)!$. Ce genre de facteurs sort de mani\`ere r\'ep\'et\'ee dans les bornes constructives et se contr\^ole tr\`es facilement par
la d\'ecroissance polynomiale des corr\'elations. En effet, on voit facilement qu'un cube $\Del^j$ de degr\'e de connectivit\'e $n^j(\Del^j)$ est connect\'e \`a au moins $n^j(\Del^j)/2$ cubes $\Del$ \`a distance $d^j(\Del,\Del^j)\ge C (n^j(\Del^j))^{1/D}$ (en raison du fait que les cubes forment un pavage r\'egulier de $\R^D$). Une partie du facteur de d\'ecroissance polynomiale
en $\frac{1}{(1+\Del^j(x,y))^r}$ peut \^etre utilis\'ee et produit l'inverse d'une factorielle locale. Si $r$ peut \^etre choisi  assez grand, les factorielles locales disparaissent.

\bigskip

\underline{3. Bornes multi-\'echelles}

\medskip

Venons-en maintenant aux vraies bornes, les pr\'ec\'edentes n'ayant qu'un objectif p\'edagogique. On fixe une \'echelle de r\'ef\'erence $j_{min}$. A priori il faudrait  consid\'erer une somme sur tous les polym\`eres
possibles d'\'echelles $\rho,\rho-1,\ldots,j_{min}$ et contenant au moins un cube \`a l'\'echelle $j$. Mais le calcul se fait comme dans la section 1 en partant de l'\'echelle $\rho$, et en descendant
au fur et \`a mesure dans les \'echelles; et la restriction du polym\`ere aux \'echelles sup\'erieures \`a une \'echelle donn\'ee n'est pas n\'ecessairement connexe. En pratique on d\'ecompose le calcul
comme suit. On fixe un ensemble fini de cubes ${\bf\Del}$ de diverses \'echelles, et on somme sur toutes les for\^ets de cluster$^*$ dont tous les cubes sont connect\'es par un ensemble de liens
horizontaux et verticaux \`a (au moins) l'un des cubes de l'ensemble $\bf\Del$. Les composantes des champs des diverses \'echelles \'etant mutuellement ind\'ependantes, on se ram\`ene \`a
calculer la contribution \`a la fonction de partition des composantes d'une \'echelle $j$ donn\'ee en consid\'erant les diff\'erentes attributions d'\'echelles possibles pour chaque champ
de chaque "vertex" $\int_{\Del} {\cal L}_{int}(\psi)(x)dx$ produit par une d\'erivation horizontale ou verticale, o\`u ${\cal L}_{int}(\psi)(x)=\prod_{i=1}^I \psi_i(x)$. Le r\'esultat est le suivant:

\BEA && \sum_{\Pi} |X_{\Pi}|\le \left( 1+\max_{k\ge j} \sup_{\Del^k\in\D^k} \left[ \sum_{\Del'\in\D^{j\to k}} \langle X_{\Del^k}X_{\Del'}\rangle \right.\right.\nonumber\\
&&\qquad \qquad \left.\left. +\sum_{k'<k} \sum_{\Del''\in \D^{j\to k'}} \langle X_{\Del^{k'}} X_{\Del''}\rangle \right] \right)^{3N},\nonumber\\ \label{eq:Xpimulti}\EEA
 o\`u $\Del^{k'}\supset\Del^k$ est l'unique cube d'\'echelle $k'<k$ contenant $\Del^k$.
 
 La somme sur $k'<k$ permet d'explorer tout le polym\`ere \`a partir de l'\'echelle $\rho$, la restriction du polym\`ere aux plus hautes \'echelles (comme nous l'avons d\'ej\`a signal\'e) pouvant
 \^etre connect\'e "par en-bas", i.e. par les lignes les plus basses.
 
 Exactement de la m\^eme mani\`ere que les lignes externes des diagrammes de Feynman multi-\'echelles  quasi-locaux de la section 1 contribuaient un "facteur de ressort"$^*$ $M^{-\beta ht_{\Gamma}}$
 assurant la convergence globale du diagramme renormalis\'e, chaque champ de bas moment $\psi^j$ produit \`a l'\'echelle $k$ (cf. \S 3.1) apporte un facteur de ressort $M^{-\beta(k-j)}$  -- ou
 plus g\'en\'eralement $M^{-\tilde{\beta}(k-j)}$ une fois d\'efalqu\'es les champs moyenn\'es \'eventuels --, de sorte que la contribution du terme entre crochets $[\ ]$  dans (\ref{eq:Xpimulti})
 est de l'ordre de 
 \BEA && \sum_{k'=j}^k M^{-\tilde{\beta}_i(k'-j)} \left[ M^{-\tilde{\beta}_{i'}(k'-j)} \sum_{\Del'\in\D^{k'}} \frac{1}{(1+d^{k'}(\Del^{k'},\Del'))^r} \right.\nonumber\\
 && \left. \qquad +\sum_{k''=j}^{k'-1} M^{-\tilde{\beta}_{i'}(k''-j)} \sum_{\Del''\in \D^{k''}} \frac{1}{(1+d^j(\Del^{k'},\Del''))^r} \right].\EEA
 La premi\`ere somme sur les cubes d'\'echelle $k'$, $\sum_{\Del'\in\D^{k'}} \frac{1}{(1+d^{k'}(\Del^{k'},\Del'))^r}$, est d'ordre 1; elle provient du d\'eveloppement en clusters horizontal
 d'\'echelle $k'$. La deuxi\`eme sur les cubes d'\'echelle $k''<k'$, $\sum_{\Del''\in \D^{k''}} \frac{1}{(1+d^j(\Del^{k'},\Del''))^r} $, est d'ordre $M^{D(k''-j)}$, elle provient
 des champs de bas moment d'\'echelle $j$, produits aux \'echelles $k'>k''>j$. Un bref calcul montre que la double somme converge si $-(\tilde{\beta}_i+\tilde{\beta}_{i'})+D<0$, condition r\'ealis\'ee
 puisque $\tilde{\beta}_i>D/2$ par d\'efinition.
 
 \medskip
 
 Il reste \`a sommer sur les champs de {\em haut} moment, puis \`a sommer sur les diff\'erents choix possibles de l'ensemble $\bf\Del$. La meilleure image possible de la proc\'edure
 est celle que donnent  les jeux de construction en bois ou en plastique pour les jeunes enfants, consistant \`a mettre les petits cubes dans les grands. Par exemple, la contribution d'un champ $\psi^j$
 de haut moment produit \`a l'\'echelle $h<j$  se calcule en consid\'erant sa restriction Res$^h_{\Del^j}\psi^j(x):=
{\bf 1}_{x\in\Del^j} \psi^j(x)$ aux petits cubes $\Del^j\subset\Del^h$ comme en \S 3.1.2. Le volume d'int\'egration d'un champ restreint vaut $M^{-D(j-h)}\cdot M^{-Dh}$, d'o\`u un facteur
de ressort $M^{-D(j-h)}$. En consid\'erant les diff\'erentes d\'ecompositions  $\psi_{i_1}^{j_1}\ldots\psi_{i_I}^{j_I}$ possibles d'un vertex en \'echelles $j_1\le \ldots\le j_I$, on
montre que la resommation des petits cubes dans les grands n'est possible que si 
\BEQ \beta_{i_I},\beta_{i_I}+\beta_{i_{I-1}},\ldots,\beta_{i_I}+\ldots+\beta_2<D, \label{eq:hyp} \EEQ
condition \'equivalente \`a l'hypoth\`ese
"high-momentum fields" de \cite{MagUnt2}, autrement dit si les champs en jeu ne sont pas trop divergents dans l'ultra-violet; dans le cas contraire, la th\'eorie est totalement instable aux
hautes \'energies. Les d\'etails peuvent \^etre trouv\'es dans \cite{MagUnt2}.

%%%%%%%%%%% 
 
\section{Deux mod\`eles}

Les sections pr\'ec\'edentes donnent l'illusion que les arguments constructifs sont totalement g\'en\'eraux et peuvent \^etre reproduits partout \`a l'identique. Dans les faits le choix des
dimensions des champs et de l'interaction, ainsi que  le flot
du groupe de renormalisation cr\'eent des situations tr\`es diff\'erentes les unes des autres, se refl\'etant notamment dans la r\'esolution des \'equations implicites donnant les param\`etres
renormalis\'es, et dans la m\'ethode de {\em domination}. Nous avons donc choisi de pr\'esenter un mod\`ele classique, le mod\`ele $\phi^4$ infra-rouge non massif en dimension $4$,
 ainsi que le cas des chemins rugueux, pour donner
au lecteur une vision moins exclusivement technique de la th\'eorie constructive.

\subsection{Th\'eorie $\phi^4$ de masse nulle}

On consid\`ere ici le champ bosonique libre non massif $\phi$  avec une interaction $\phi^4$ en dimension 4; c'est ce qu'on appelle habituellement la {\em th\'eorie $\phi^4$}. Il est bien connu que la constante de couplage diverge logarithmiquement dans la limite {\em ultra-violette}.
La divergence ultra-violette de ce mod\`ele est similaire \`a celle de l'\'electrodynamique quantique, d'o\`u son utilisation fr\'equente comme "mod\`ele jouet" (toy model) pour comprendre les difficult\'es
de la renormalisation en physique des hautes \'energies \footnote{Notons n\'eanmoins que le boson de Higgs massif (cl\'e de vo\^ute du mod\`ele standard mais non encore observ\'e) rentre dans le
cadre de la th\'eorie $\phi^4$.}. Au contraire, la classification par Landau et Ginzburg des transitions de phases du second ordre \cite{LeB} sugg\`ere que le {\em mod\`ele $\phi^4$ \em infra-rouge}
et le {\em mod\`ele d'Ising} sont dans la m\^eme classe d'universalit\'e; la masse du champ bosonique est proportionnelle \`a l'\'ecart de la temp\'erature \`a la temp\'erature critique, et le
champ non massif refl\`ete donc  le mod\`ele d'Ising pr\'ecis\'ement \`a la temp\'erature critique \footnote{La fonction de Green $G(x,y)$ de l'op\'erateur $\Del+m^2$ se comportant comme
$e^{-m|x-y|}$ \`a grande distance, $1/m$ joue le r\^ole d'une {\em longueur de corr\'elation}, infinie \`a la temp\'erature critique.}. Du c\^ot\'e infra-rouge au contraire, la th\'eorie
devient asymptotiquement libre, la constante de couplage ayant une d\'ecroissance logarithmique; on prouve en fait que les fonctions  \`a $n$ points sont celles de la th\'eorie
libre non massive, \`a des corrections logarithmiques pr\`es.
 
Rentrons maintenant dans les d\'etails. Le  lagrangien d'interaction {\em nu} de ce mod\`ele s'\'ecrit
${\cal L}_{int}(\phi)=\lambda^0 (\phi^{\to 0}(x))^4$, o\`u $\phi^{\to 0}=\sum_{j=-\infty}^0 \phi^j$ est un champ gaussien sur $\R^4$ de covariance $\langle \phi^{\to 0}(x)\phi^{\to 0}(y)\rangle=
{\cal F}^{-1}\left( \frac{\chi^{\to 0}(\xi)}{|\xi|^2}\right)(x-y)$, correspondant approximativement \`a la troncature ultra-violette \`a l'\'echelle $j=0$ d'une mesure gaussienne
qu'on pourrait \'ecrire (de mani\`ere impropre) $e^{-\half \int_{\R^4} |\nabla\phi|^2(x) dx} {\cal D}\phi$ \footnote{Dans l'interpr\'etation de ce mod\`ele en physique statistique, ce cut-off ultra-violet
revient \`a consid\'erer le mod\`ele d'Ising sur un r\'eseau de maille de longueur $a=1$ dans les unit\'es choisies.}. Les d\'eveloppements en clusters conduisent \`a un lagrangien habill\'e$^*$
\BEA && {\cal L}_{int}(\phi;\vec{t})(x)=\lambda^{0} \left( (T\phi)^{\to 0}\right)^4(x)+\sum_{\rho'\le 0} \lambda^{\rho'-1} (1-(t_x^{\rho'})^4) \left( (T\phi)^{\to(\rho'-1)}\right)^4(x) \nonumber\\
&& \qquad \qquad  \qquad \qquad +
\del {\cal L}_{masse}(\phi;t)(x)+\del {\cal L}_{onde}(\phi;t)(x),\EEA
o\`u 
\BEQ \del {\cal L}_{masse}(\phi;\vec{t})(x)=\sum_{\rho'\le 0} (\del m^2)^{\rho'-1} (1-(t_x^{\rho'})^2) \left( (T\phi)^{\to(\rho'-1)}\right)^2(x) \label{eq:4.2} \EEQ
est le {\em contre-terme de masse}, et
\BEQ \del {\cal L}_{onde}(\phi;\vec{t})(x)=\sum_{\rho'\le 0} (\del Z_3)^{\rho'-1} (1-(t_x^{\rho'})^2) \left| \nabla (T\phi)^{\to(\rho'-1)}\right|^2(x) \EEQ
le {\em contre-terme de fonction d'onde}. Au risque de nous r\'ep\'eter, ${\cal L}_{int}(\phi;1)={\cal L}_{int}(\phi)$ est le lagrangien initial de la th\'eorie, et l'habillage avec
les param\`etres $t$, resp. les contre-termes $\del {\cal L}_{masse}$ et $\del {\cal L}_{onde}$ ne proviennent que de d\'eveloppements d'ordre combinatoire et de resommations partielles
de l'interaction; il ne s'agit donc pas de termes suppl\'ementaires rajout\'es \`a la main pour supprimer les divergences, comme on l'entend dire parfois.

\begin{figure}[h]
  \centering
   \includegraphics[scale=0.5]{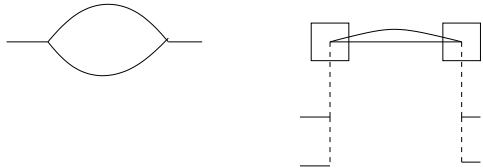}
   \caption{Diagramme bulle.}
  \label{Fig-phi41}
\end{figure}

Des calculs perturbatifs initiaux donnent une bonne id\'ee du flot de la constante de couplage $\lambda$ et des contre-termes $\del m^2$, $\del Z_3$. On s'inspirera ici des \'equations
de Callan-Symanzik donnant le flot du  groupe de renormalisation, telles que d\'ecrites dans le chapitre 7 du livre de M. Le Bellac \cite{LeB}. La constante de couplage flottante $\lambda(\xi)$
 y est d\'efinie comme la somme des contributions de tous les diagrammes "one-particle irreducible" (1 P.I.) avec $4$ propagateurs externes de moments d'ordre
de grandeur $\xi$ (cf. \cite{LeB}, \'eq. (7.1.3c)). Avec notre d\'ecoupage en \'echelles, il est naturel ici de remplacer $\lambda(M^{-j})$  par $\lambda^{-j}$, qui est formellement la somme
sur tous les diagrammes de lignes internes d'\'echelles $\ge -j$. La fonction
$\beta(\lambda)=\frac{d\lambda}{d\ln\xi}$ (calcul\'ee \`a $\lambda^0$ -- constante {\em nue}$^*$ -- fix\'ee), donnant le flot de la constante de couplage, est approch\'ee ici par le flot discret
$\lambda^{-j-1}-\lambda^{-j}=-\beta(\lambda^{-j})$. La contribution principale \`a la fonction $\beta$ provient du diagramme bulle de la figure \ref{Fig-phi41} \'evalu\'ee \`a moments
externes nuls, avec la contrainte que l'un au moins des propagateurs (donc les deux par conservation des moments) soit d'\'echelle $-j$, ce qui donne 
\BEQ \beta(\lambda^{-j})\approx (\lambda^{-j})^2 \int_{M^{-j-1}<|\xi|<M^{-j}} \frac{d^4\xi}{(|\xi|^2)^2} + O((\lambda^{-j})^3)= c(\lambda^{-j})^2+
O((\lambda^{-j})^3), \EEQ
 o\`u $c>0$ est une constante.

\begin{figure}[h]
  \centering
   \includegraphics[scale=0.5]{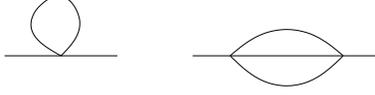}
   \caption{De gauche \`a droite: diagramme ``tadpole'' (t\^etard); diagramme \`a deux boucles.}
  \label{Fig-phi42}
\end{figure}

 Le syst\`eme dynamique discret $\lambda^{-j-1}-\lambda^{-j}=-c(\lambda^{-j})^2$ se comporte asymptotiquement comme le syst\`eme
dynamique continu $\frac{d\lambda}{ds}=-c\lambda^2$, avec $s:=-\ln\xi$, donnant $\lambda^{-j}\simeq \frac{1}{(1/\lambda^0)+cj}\sim_{j\to\infty} \frac{1}{cj}$.
Les contre-termes $(\del m^2)^{-j}$ et $(\del Z_3)^{-j}$ se calculent de la m\^eme mani\`ere en consid\'erant le d\'eveloppement de Taylor \`a l'ordre 2 du diagramme "t\^etard" (tadpole),  d'ordre
$\lambda$ (qui contribue uniquement au contre-terme de masse),
et du diagramme \`a deux boucles, d'ordre $\lambda^2$ (cf. Fig. \ref{Fig-phi42}. On trouve pour le diagramme "tadpole"
\BEQ (\del m^2)^{-j-1}-(\del m^2)^{-j}\approx \lambda^{-j}\int_{M^{-j-1}<|\xi|<M^{-j}} \frac{d^4\xi}{\xi^2}=c_m\lambda^{-j}M^{-2j}\sim_{j\to\infty} \frac{M^{-2j}}{cj},\EEQ
et pour l'autre diagramme une nouvelle contribution \`a $(\del m^2)^{-j-1}-(\del m^2)^{-j}$ de l'ordre de $(\lambda^{-j})^2 M^{-2j}$, ainsi que
\BEQ (\del Z_3)^{-j-1}-(\del Z_3)^{-j}\approx c_Z(\lambda^{-j})^2\sim_{j\to\infty} \frac{c_Z}{c^2 j^2}.\EEQ
Au lieu d'une condition "initiale" (\`a l'\'echelle 0), on fixe une condition "terminale" $(\del m^2)^{-\infty}=0$, d\'ecrivant un champ non massif (cf. note 9 en bas de page 
\pageref{eq:4.2}), et de m\^eme $(\del Z_3)^{-\infty}=0$. 
On trouve alors imm\'ediatement $(\del m^2)^{-j}=O(M^{-2j}/j)$, et $(\del Z_3)^{-j}=O(1/j)$ {\em fini}, gr\^ace \`a la
convergence de la s\'erie $\sum_{k\ge 1} \frac{1}{k^2}$. 

Ces r\'esultats sont confirm\'es par les bornes constructives de la section pr\'ec\'edente, en rempla\c cant les sommes formelles sur les diagrammes 1 P.I. par la somme (convergente \`a chaque
\'echelle) sur les polym\`eres  d'\'echelle minimum $-j$ \`a deux ou quatre propagateurs externes. Comme dans la th\'eorie perturbative, les parties locales des polym\`eres divergents sont resomm\'ees
en les termes suppl\'ementaires dans le lagrangien. Le champ $\phi$ est de dimension d'\'echelle $D/2-1=1$, l'hypoth\`ese "high-momentum fields"  (cf. \'eq. (\ref{eq:hyp})) est donc
v\'erifi\'ee. En revanche $1<D/2=2$, ce qui impose (cf. \S 3.1) de s\'eparer les champs moyenn\'es de bas moment $\phi^{\to j}(\Del^j)$. Cette s\'eparation fait que ceux-ci ne sont plus compens\'es par les
contre-termes; on doit donc s\'eparer les termes analogues dans les contre-termes \'egalement. Tous ces termes doivent \^etre domin\'es par l'exponentielle de l'interaction. Celle-ci est
inf\'erieure \`a $\exp \left(-\lambda^0 \int ((T\phi)^{\to 0})^4 (x) dx -\sum_{\rho'\le 0} \lambda^{\rho'-1} (1-(t_x^{\rho'})^4 )\int \left( (T\phi)^{\to(\rho'-1)}\right)^4(x) dx \right)$.
Les champs moyenn\'es de bas moment sont produits en faisant agir un op\'erateur de d\'erivation horizontal ou vertical, $\Hor$ ou $\Vert$, sur un vertex $\int_{\Del^k} \phi^4(x) dx$. Trois
sur quatre tout au plus de ces champs prennent l'ascenseur ensemble et descendent (s\'epar\'ement ou pas) \`a des  \'etages  $j$, qui sont autant d'\'echelles de "l\^achage"$\ ^*$. On peut
accompagner chacun d'un facteur $\lambda^{\kappa}$, o\`u $1/4<\kappa<1/3$, de sorte que les autres champs du vertex laiss\'es \`a la porte de l'ascenseur puissent se partager un {\em petit
facteur} $\lambda^{\kappa'}$, $\kappa'>0$ servant aux bornes gaussiennes. Consid\'erons tous les champs moyenn\'es de bas moment $\lambda^{\kappa} \phi^{\to -j}(\Del^{-j})$ l\^ach\'es dans un cube $\Del^{-j}$. On fait
maintenant la remarque simple suivante, refl\'etant la {\em stricte positivit\'e} du lagrangien (cf. remarque dans l'introduction \`a la section 3):
\BEQ \lambda^{\kappa} |\phi^{\to -j}(\Del^{-j})| \exp -\lambda \int_{\Del^{-j}} (\phi^{\to -j})^4(x) dx\le \frac{\lambda^{\kappa-1/4}}{|\Del^{-j}|},\EEQ
cons\'equence de l'in\'egalit\'e triviale $|x|e^{-A|x|}=A^{-1} (A|x|e^{-A|x|})\le A^{-1}$. En v\'erit\'e, il faudrait tenir compte des deux faits suivants: (i) les champs $\phi$ (champs moyenn\'es comme
ceux pr\'esents dans l'interaction) sont d\'ecor\'es de $t$-facteurs $1-t^{-j}_{\Del^{-j}}$ ou $1-(t^{-j}_{\Del^{-j}})^4$; (ii) le cube peut contenir jusqu'\`a
 $O(n(\Del^{-j}))$ champs moyenn\'es ($n(\Del^{-j})$ \'etant le degr\'e de 
connectivit\'e de $\Del^{-j}$).  On montre facilement qu'ils ne changent rien au calcul pr\'ec\'edent.

%%%%%%%%%%%%%%%%%%%%%

\subsection{Chemins rugueux}

\begin{figure}[h]
  \centering
   \includegraphics[scale=0.5]{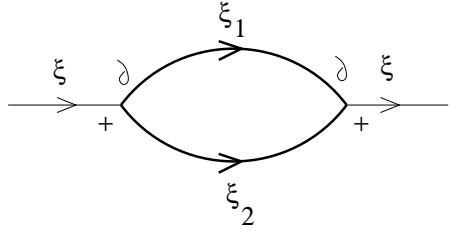}
   \caption{Diagramme bulle.}
  \label{Fig-bubble}
\end{figure}

Ce n'est pas le lieu d'expliquer en d\'etails la th\'eorie des  chemins rugueux, ainsi que l'origine de la solution que nous avons apport\'ee au probl\`eme de la d\'efinition de l'aire de L\'evy du brownien
fractionnaire. Le lecteur int\'eress\'e pourra se r\'ef\'erer \`a \cite{Unt-Holder,Unt-fBm,Unt-ren,FoiUnt}. Les constructions explicites de chemins rugueux que nous avons introduites reposent
toutes sur des m\'ethodes multi-\'echelles, et plus particuli\`erement sur la {\em mise en ordre normal de Fourier}$^*$ des {\em int\'egrales squelette}$^*$. D\'efinissons bri\`evement ces notions
dans le cas du brownien fractionnaire et des int\'egrales it\'er\'ees d'ordre 2. On consid\`ere un champ stationnaire gaussien $\phi=(\phi_1,\phi_2)$ sur $\R$ (ici $D=1$) \`a deux composantes ind\'ependantes, de covariance Fourier $\langle |\hat{\phi}_i(\xi)|^2\rangle=\frac{1}{|\xi|^{1+2\alpha}}$, o\`u $\alpha\in]0,\half[$ est l'indice de Hurst du brownien fractionnaire.  Ce champ poss\`ede une divergence {\em infra-rouge} due \`a la non-int\'egrabilit\'e
de ce noyau en $\xi=0$; le brownien fractionnaire $B(t):=\phi(t)-\phi(0)$ est bien d\'efini en revanche \footnote{Sa covariance dans l'espace direct s'\'ecrit $\langle B_s B_t\rangle=
\half(|t|^{2\alpha}+|s|^{2\alpha}-|t-s|^{2\alpha})$.}. De mani\`ere g\'en\'erale les divergences infra-rouge qui apparaissent dans les calculs
interm\'ediaires disparaissent lorsqu'on consid\`ere des incr\'ements, ce qui est le cas de toutes les quantit\'es construites \`a partir du   brownien fractionnaire, et on n'y fera plus
attention. On d\'ecompose le champ $\phi$ en \'echelles $\phi^j$ comme pr\'ec\'edemment. Contrairement au cas de la th\'eorie $\phi^4$ infra-rouge abord\'ee dans le paragraphe pr\'ec\'edent,
c'est le comportement ultra-violet  aux  grandes \'echelles $j\ge 0, j\to +\infty$ qui demande une attention particuli\`ere. Remarquons tout d'abord que l'{\em aire de L\'evy} de $B$, ${\cal A}(s,t)=\int_s^t dB_1(t_1)\int_s^{t_1} dB_2(t_2)$, se d\'ecompose successivement en somme de plusieurs termes,
\BEQ {\cal A}(s,t)={\cal P}^+ {\cal A}(s,t)+{\cal P}^- {\cal A}(s,t),\EEQ
o\`u le projecteur de Fourier ${\cal P}^{\pm}$ envoie $\phi_1\otimes \phi_2$ sur $\half\sum_{j} \phi_1^j\otimes \phi_2^j+\sum_{j\lessgtr k}\phi_1^{j}\otimes\phi_2^k$; puis
\BEQ {\cal P}^+ {\cal A}(s,t)= {\cal P}^+ \int_s^t dB_1(t_1)\int^{t_1} dB_2(t_2)- {\cal P}^+ \int_s^t dB_1(t_1) \int^s dB_2(t_2),\EEQ
les int\'egrales $\int^{t_1}$ ou $\int^s$ \'etant des {\em int\'egrales squelette} (donn\'ees formellement par la multiplication par $\frac{1}{\II\xi}$ en  Fourier). On montre que
le deuxi\`eme terme ${\cal P}^+ \int_s^t dB_1(t_1)\int^s dB_2(t_2)={\cal P}^+ (\phi_1(t)-\phi_1(s))\phi_2(s)$ est $2\alpha-\eps$-H\"older pour tout $\eps>0$. La partie singuli\`ere
de l'aire de L\'evy (singuli\`ere pour $\alpha\le 1/4$ en tout cas) est donc donn\'ee par  les incr\'ements de deux fonctions ${\cal A}^{\pm}$, de d\'eriv\'ee
\BEQ \partial{\cal A}^+(t):={\cal P}^+ \partial \phi_1(t)\phi_2(t)=\half \sum_j \partial\phi_1^j(t) \phi_2^j(t)+\sum_{j<k} \partial\phi_1^j(t)\phi_2^k(t)\EEQ
et similairement pour $\partial{\cal A}^-$ en \'echangeant les indices, de sorte que l'\'echelle du champ d\'eriv\'e $\partial\phi$ soit inf\'erieure \`a celle du champ
non d\'eriv\'e.
On v\'erifie facilement que la transform\'ee de Fourier de la fonction \`a deux points $\langle \partial{\cal A}^{\pm}(s)\partial{\cal A}^{\pm}(t)\rangle$ 
est donn\'ee par le diagramme de Feynman "bulle" amput\'e de la figure \ref{Fig-bubble}, \'egal \`a peu de choses pr\`es \`a $\int_{|\xi_1|<|\xi-\xi_1|} \frac{d\xi_1}{|\xi_1|^{-1+2\alpha}
|\xi-\xi_1|^{1+2\alpha}}\approx \int^{+\infty} \frac{d\xi}{\xi^{4\alpha}}$, et diverge donc si et seulement si $\alpha\le 1/4$. On retrouve ainsi rapidement les r\'esultats classiques
de L. Coutin et Z. Qian \cite{CQ02}. 

Supposons dor\'enavant $\alpha<1/4$. Pour aller plus loin,  il est naturel de consid\'erer les "wavy lines" ext\'erieures comme propagateurs d'un champ gaussien $\sigma=(\sigma_+,\sigma_-)$ 
\`a deux composantes ind\'ependantes, qu'on choisit de covariance
Fourier $\langle |\hat{\sigma}_{\pm}(\xi)|^2\rangle=\frac{1}{|\xi|^{1-4\alpha}}$ par homog\'en\'eit\'e. On est donc conduit \`a introduire le lagrangien d'interaction
\BEQ {\cal L}_{int}(x)=\II\lambda \left(\partial{\cal A}^+(x)\sigma_+(x)+\partial{\cal A}^-(x)\sigma_-(x)\right).\EEQ
Les dimensions des champs sont $-\alpha$ pour $\phi$, $2\alpha$ pour $\sigma$, et la constante de couplage $\lambda\in\R,\lambda\not=0$ est  sans dimension, de sorte que la
th\'eorie est a priori juste renormalisable. Les conditions tr\`es particuli\`eres sur les \'echelles dans les vertex (le champ d\'eriv\'e $\partial\phi$ \'etant de {\em bas moment} par rapport au
champ $\phi$) impliquent que $\lambda$ n'est pas renormalis\'ee \footnote{En effet, les diagrammes de Feynman avec trois propagateurs externes $\phi,\partial\phi,\sigma$ ne peuvent
\^etre des diagrammes dangereux au sens de la section 1.}. Le comptage de puissance et les sym\'etries de la th\'eorie montrent que seuls les diagrammes \`a $2n$ propagateurs externes $\sigma$
et $0$ propagateur externe $\phi$ sont potentiellement divergents, de degr\'es de divergence  $\omega_{2n}=1-4n\alpha$. Pour $\alpha\in]1/4,1/8[$ (hypoth\`ese \`a laquelle nous nous tenons
par la suite) seul le propagateur du champ $\sigma$ (correspondant \`a $n=1$) doit \^etre renormalis\'e. 

Des arguments perturbatifs na\"\i fs font bien comprendre ce qui se passe. Le diagramme "bulle" diverge comme $(\II\lambda)^2 M^{\rho(1-4\alpha)}$ pour un cut-off ultraviolet de l'ordre
de $M^{\rho}$. La s\'erie g\'eom\'etrique associ\'ee \`a ce diagramme 1 P.I. donne pour le propagateur renormalis\'e du champ $\sigma$
$$ \frac{1}{|\xi|^{1-4\alpha}}\left( 1-\lambda^2\frac{M^{\rho(1-4\alpha)}}{|\xi|^{1-4\alpha}}+\ldots\right)=\frac{1}{|\xi|^{1-4\alpha}+\lambda^2 M^{\rho(1-4\alpha)}},$$
tendant exponentiellement vite vers $0$ quand $\rho\to\infty$. Le champ $\sigma$ acquiert donc une masse $\del m$ de l'ordre de $M^{\rho(1-4\alpha)}$. Autrement dit, l'interaction dispara\^\i t \`a toutes les \'echelles lorsqu'on supprime le cut-off! Le champ $\phi$ reste donc
gaussien, avec exactement la m\^eme fonction de covariance initiale. Cependant, la m\^eme s\'erie donne une covariance renormalis\'ee 
$\langle \left|{\cal F}\partial{\cal A}^+(\xi)\right|^2\rangle$ {\em finie}, \'egale \`a $\frac{M^{\rho(1-4\alpha)}}{1+\lambda^2 (M^{\rho}/|\xi|)^{1-4\alpha}}\to_{\rho\to\infty}
\frac{|\xi|^{1-4\alpha}}{\lambda^2}$, et redonnant apr\`es transformation de Fourier inverse une aire de L\'evy de r\'egularit\'e H\"older $2\alpha-\eps$. 

Les arguments constructifs font appara\^itre une subtilit\'e (insoup\c connable avec des arguments purement perturbatifs), li\'ee au probl\`eme de la domination des champs $\sigma$
moyenn\'es de bas moment. Soit $b^j:=((\del m)^{j-1}-(\del m)^{j})/\lambda^2$ le contre-terme de masse d'\'echelle $j$,
divis\'e par le carr\'e de la constante de couplage.  Par un raisonnement similaire \`a celui utilis\'e pour  la th\'eorie $\phi^4$, chaque champ $\sigma$ de bas moment produit \`a partir
du vertex en $\lambda (\partial{\cal A})\sigma$ peut prendre l'ascenseur accompagn\'e d'un petit facteur $O(\lambda^{\kappa})$,
o\`u $\kappa<1$. A priori il ne peut \^etre domin\'e que par le contre-terme de masse en $\lambda^2 b^j \sigma^2$. Sch\'ematiquement $\lambda^{\kappa}\sigma e^{-\lambda^2 \sigma^2}=
  O(\lambda^{\kappa-1})$, un {\em grand facteur} au lieu du {\em petit facteur} souhait\'e. Afin de r\'esoudre ce probl\`eme on est amen\'e \`a rajouter un {\em terme de bord}$^*$ dans
  l'interaction,  qu'on peut choisir \'egal \`a ${\cal L}_{12}:=M^{-(12\alpha-1)\rho} \lambda^3 (\sigma(x))^6$. Le pr\'efacteur $M^{-(12\alpha-1)\rho}$ a \'et\'e choisi de mani\`ere \`a ce que
   l'int\'egrale de $M^{-(12\alpha-1)\rho} (\sigma^{\rho})^6$ sur un intervalle $\Del^{\rho}$ soit de l'ordre de $1$. Comme $12\alpha-1>0$, ce terme est en fait \'evanescent \`a toutes 
   les \'echelles $j\ll \rho$.  A ces \'echelles, la domination par le contre-terme $\lambda^2 b\sigma^2$ produit un facteur $1$, multipli\'e par un petit facteur $M^{-\half(1-4\alpha)(\rho-j)}$ d\^u \`a $b^j\sim M^{(1-4\alpha)j}$.
   Si au contraire $j\simeq\rho$ ("\'echelles de bord"), ce facteur vaut \`a peu pr\`es $1$, et on domine alors par le terme de bord ${\cal L}_{12}$. Sch\'ematiquement,
   $\lambda \sigma^j e^{-M^{-(12\alpha-1)\rho}\lambda^3 (\sigma^j)^6}=O(\lambda^{\half})$, que multiplie $M^{(12\alpha-1)(\rho-j)/6}$. En choissant $\lambda$ assez petit on arrive \`a obtenir
   un petit facteur \`a toutes les \'echelles. Il faut encore v\'erifier qu'on peut dominer les champs $\sigma$ moyenn\'es de bas moment issus de ${\cal L}_{12}$ par 
   ${\cal L}_{12}$ lui-m\^eme.  Les d\'etails peuvent \^etre trouv\'es dans \cite{MagUnt2}.

\newpage
%%%%%%%%%%%%%%%%%%%%

{\huge \bf Lexique fran\c cais-anglais}

\medskip

A l'usage des anglophones et de ceux qui n'ont jamais connu que les termes anglais, il se conclut par quelques termes sp\'ecifiques aux chemins rugueux, utilis\'es dans le dernier
paragraphe de l'article uniquement.

\medskip

\begin{tabular}{ll}

arbre couvrant, {\em spanning tree} & arbre enracin\'e, {\em rooted tree} \\

arbres de polym\`eres, {\em polymer trees} &
bas moment, {\em low-momentum} \\

bornes gaussiennes, {\em Gaussian bounds} &
  boucle, {\em loop} \\

champs moyenn\'es de bas moment, & champs restreints de haut moment, \\ 
 \quad {\em averaged low-momentum fields} & \quad {\em restricted high-momentum fields}\\

champ secondaire, {\em secondary field} &  chemin rugueux, {\em rough path} \\

comptage de puissance, 
  {\em power-counting}  &  constante nue, {\em bare constant} \\
  
   constante renormalis\'ee, {\em renormalized constant} &
 
degr\'e de libert\'e, {\em degree of freedom} \\

d\'eveloppement en clusters horizontal, & d\'eveloppement en clusters vertical, \\ 
\quad  {\em horizontal cluster expansion}, & \quad {\em vertical cluster expansion} \\

diagramme du vide, {\em vacuum diagram} & \'echelle de chute libre, {\em free falling scale} \\

\'echelle de derni\`ere $t$ d\'erivation cons\'ecutive  & \'echelle de production, \\
   \quad ou de l\^achage, {\em dropping scale} & \quad {\em production scale} \\

 \'echelle propre, {\em proper scale} & factorielles locales, {\em local factorials} \\

fonction d'\'evaluation du polym\`ere,  & fusionner, {\em merge} \\
 \quad {\em polymer evaluation function} &  \\

haut moment, {\em high-momentum} & \^\i lots horizontaux, {\em horizontal islands} \\

lagrangien habill\'e, {\em dressed Lagrangian} & lien d'inclusion, {\em inclusion link} \\

op\'erateur de d\'eveloppement horizontal,  & param\`etre d'affaiblissement, 
\\  \quad {\em horizontal expansion operator} & \quad {\em weakening parameter} \\

 polym\`ere du vide,
 {\em vacuum polymer} &  ressort, {\em spring} \\

troncature, {\em cut-off} & \\

& \\
& \\

aire de L\'evy, {\em L\'evy area} & brownien fractionnaire, \\
int\'egrale squelette, {\em skeleton integral} & \quad {\em fractionary Brownian motion} \\
mise en ordre normal de Fourier,  & terme de bord,\\  \quad {\em Fourier normal ordering} & \quad {\em boundary term}

\end{tabular}

%%%%%%%%%%%%%%%%%%%%%%%%%%%%%%%%%%%%%%%%%%%
%%%%%%%%%%%%%%%%%%%%%%%%%%%%%%%%%%%%%%%%%%%

%%%%%%%%%%%%%%%%%%%%%%%%%%%%%%%%%%
%%%%%%%%%%%%%%%%%%%%%%%%%%%%%%%%\`u
%%%%%%%%%%%%%%%%%%%%%%\`u\`u
%%%%%%%%%%%%%%%%%%%%%%%
%%%%%%%%%%%%%%%%%%%%%%%%%
%%%%%%%%%%%%%%%%%%%%%%%
\end{document}